\documentclass[journal]{IEEEtran}

\IEEEoverridecommandlockouts

\usepackage[cmex10]{amsmath}
\usepackage{amssymb}

\usepackage{cases}
\usepackage{multirow}
\usepackage{empheq}

\usepackage{units}
\usepackage[noadjust]{cite}
\usepackage{verbatim}

\usepackage{algorithm}
\usepackage{algpseudocode}

\usepackage{siunitx}

\ifCLASSINFOpdf
\usepackage[pdftex]{graphicx}
\else
\usepackage[dvips]{graphicx}
\fi
\usepackage[caption=false,font=footnotesize]{subfig}

\usepackage{color}
\usepackage{multirow}

\usepackage[hyphens]{url}

\newtheorem{theorem}{Theorem}
\newtheorem{lemma}[theorem]{Lemma}

\newtheorem{corollary}[theorem]{Corollary}
\newtheorem{definition}[theorem]{Definition}

\newtheorem{remark}[theorem]{Remark}

\newcommand{\vect}[1]{\mathbf{#1}}

\begin{document}
	
	\sloppy
	
	\title{On the Efficient Estimation of Min-Entropy}
	
	\author{
		\IEEEauthorblockN{Yongjune Kim, Cyril Guyot, and Young-Sik Kim}	\\		
		
		\thanks{Y. Kim is with the Department of Information and Communication Engineering, DGIST, Daegu 42988, South Korea (e-mail: yjk@dgist.ac.kr). C. Guyot is with Western Digital Research, Milpitas, CA 95035 USA (e-mail: cyril.guyot@wdc.com). Y.-S. Kim is with the Department of Information and Communication
		Engineering, Chosun University, Gwangju 61452, South Korea (e-mail: iamyskim@chosun.ac.kr). Corresponding authors: Yongjune Kim and Young-Sik Kim}		
	}

	
	\maketitle
	
	\begin{abstract}
		The min-entropy is a widely used metric to quantify the randomness of generated random numbers in cryptographic applications; it measures the difficulty of guessing the most likely output. An important min-entropy estimator is the \emph{compression estimator} of NIST Special Publication (SP) 800-90B, which relies on Maurer's universal test. In this paper, we propose two kinds of min-entropy estimators to improve computational complexity and estimation accuracy by leveraging two variations of Maurer's test: Coron's test (for Shannon entropy) and Kim's test (for R{\'{e}}nyi entropy). First, we propose a min-entropy estimator based on Coron's test. It is computationally more efficient than the compression estimator while maintaining the estimation accuracy. The secondly proposed estimator relies on Kim's test that computes the R{\'{e}}nyi entropy. This estimator improves estimation accuracy as well as computational complexity. We analytically characterize the bias-variance tradeoff, which depends on the order of R{\'{e}}nyi entropy. By taking into account this tradeoff, we observe that the order of two is a proper assignment and focus on the min-entropy estimation based on the collision entropy (i.e., R{\'{e}}nyi entropy of order two). The min-entropy estimation from the collision entropy can be described by a closed-form solution, whereas both the compression estimator and the proposed estimator based on Coron's test do not have closed-form solutions. By leveraging the closed-form solution, we also propose a lightweight estimator that processes data samples in an online manner. Numerical evaluations demonstrate that the first proposed estimator achieves the same accuracy as the compression estimator with much less computation. The proposed estimator based on the collision entropy can even improve the accuracy and reduce the computational complexity.   
	\end{abstract}
	
	\section{Introduction}
	
	Random numbers are essential for generating cryptographic information such as secret keys, nonces, salt values, \emph{etc}. The security of cryptographic systems crucially depends on the randomness of generated random numbers~\cite{Turan2018recommendation,Hagerty2012entropy,Kelsey2015predictive,Amaki2013worst,Ma2019entropy}. Randomness of generated numbers should be quantified and \emph{entropies} are widely used metrics as in AIS.31~\cite{Killmann2011proposal}, NIST Special Publication (SP) 800-22~\cite{Rukhin2010statistical}, and NIST SP 800-90B~\cite{Turan2018recommendation}.
	
	There are several kinds of entropies such as Shannon entropy, R{\'{e}}nyi entropy, and min-entropy. The Shannon entropy quantifies the difficulty of guessing a typical output of random sources~\cite{Hagerty2012entropy}, which was used in AIS.31~\cite{Killmann2011proposal} and NIST SP 800-22~\cite{Rukhin2010statistical}. The min-entropy corresponds to the difficulty of guessing the most likely output of random sources~\cite{Turan2018recommendation,Kelsey2015predictive}. The NIST SP 800-90B~\cite{Turan2018recommendation} supports the use of min-entropy to quantify the randomness. 
	
	For independent and identically distributed (IID) sources, the empirical estimator (or its bias-corrected estimator) can efficiently estimate the Shannon entropy~\cite{Paninski2004estimating} and the R{\'{e}}nyi entropy~\cite{Acharya2015complexity}. The NIST SP 800-90B determines the min-entropy estimates of IID sources by using the empirical estimator (i.e., \emph{most common value estimator} of NIST SP 800-90B).

	However, it is difficult to estimate the min-entropy of non-IID sources accurately~\cite{Killmann2011proposal}. Hence, the NIST SP 800-90B adopted ten different algorithms to estimate the min-entropy~\cite[Ch 6.3]{Turan2018recommendation}. These estimators independently perform their own estimations based on different statistics. Then, the NIST SP 800-90B conservatively selects the minimum among these ten different estimates as a final estimate of min-entropy. 
	
	Although this conservative approach is preferred in security applications, it incurs a problem of \emph{detrimental underestimate}. Even if only one estimator provides a significant underestimate, it determines the final estimation no matter how accurate the other estimators are~\cite{Kelsey2015predictive,Zhu2017analysis,Zhu2020analysis}. Hence, it is important to avoid significant underestimates to obtain more accurate min-entropy estimates. 
	
	In this paper, we focus on improving the \emph{compression estimator} among ten min-entropy estimators of the NIST SP 800-90B (see Table~\ref{tab:estimators}) since it often underestimates the min-entropy. The compression estimator theoretically relies on Maurer's universal test~\cite{Maurer1992universal}. Maurer's test quantifies randomness by taking into account the minimum distances between matching patterns. It is a computationally efficient algorithm whose output is closely related to the Shannon entropy~\cite{Maurer1992universal}. Maurer's test is a widely used randomness test; it was included in the NIST SP 800-22~\cite{Rukhin2010statistical}. By leveraging Maurer's test, Hagerty and Draper~\cite{Hagerty2012entropy} proposed an algorithm to estimate the lower bound on the min-entropy. Afterward it became the compression estimator of the NIST SP 800-90B. To the best of our knowledge, the compression estimator is the only method to estimate the min-entropy by using the minimum distances between matching patterns.
	
	Although the compression estimator is theoretically well-justified by Maurer's test, it is prone to underestimate the min-entropy as discussed in~\cite{Hagerty2012entropy,Kelsey2015predictive,Zhu2017analysis}. The reasons for underestimation are twofold: 
	\begin{itemize}
		\item \emph{Bias}: Since the compression estimator estimates the lower bound on the min-entropy instead of the actual min-entropy~\cite{Hagerty2012entropy}, the bias is inevitable. Depending on the distributions of sources, the bias between the estimated min-entropy and the actual min-entropy can be large~\cite{Hagerty2012entropy,Kelsey2015predictive,Zhu2017analysis}.         
		\item \emph{Variance}: To ensure the confidence level of \unit[99]{\%}, the lower bound of the confidence interval for the Maurer's test value is used to estimate the min-entropy. Hence, the larger variance of Maurer's tests leads to the lower underestimate of min-entropy.      		
	\end{itemize}
	
	The impact of variance of Maurer's test can be reduced by including more samples. However, the computational complexity of the compression estimator is $\mathcal{O}(MK)$ (where $K$ denotes the number of samples and $M$ denotes the number of iteration for the bisection method), which limits the improvement of estimation by including more samples. 
	
	We propose two types of computationally efficient min-entropy estimators. First, we propose a min-entropy estimator by using Coron's test~\cite{Coron1999on} instead of Maurer's test. Its computational complexity is $\max\{\mathcal{O}(K), \mathcal{O}(M)\}$ instead of $\mathcal{O}(MK)$, so we can efficiently include more data samples to reduce the impact of variance. 
	
	The proposed estimator based on Coron's test is motivated by observing that the compression estimator's approach~\cite{Hagerty2012entropy} is essentially similar to the approach in~\cite{Tebbe1968uncertainty,Golic1987relationship,Feder1994relations}, which relates the lower bound on the probability of error and the Shannon entropy. The proposed estimator based on Coron's test is more efficient than the compression estimator while achieving the same accuracy as the compression estimator. We note that the proposed estimator based on Coron's test would be an appealing alternative to the compression estimator of the NIST SP 800-90B. 
	 
	In spite of the computational advantage of the min-entropy estimator based on Coron's test, it does not reduce the bias. It is because the test values by Maurer's test and Coron's test are inherently similar~\cite{Maurer1992universal,Coron1999on}. We propose a min-entropy estimator based on the R{\'{e}}nyi entropy to reduce the bias.  
	
	Recently, Kim~\cite{Kim2018low} proposed a variation of Maurer's test to estimate the R{\'{e}}nyi entropy. By leveraging Kim's test, we propose a min-entropy estimator that effectively reduces the bias compared to the compression estimator. We show that the bias can be decreased by adopting a higher order of the R{\'{e}}nyi entropy. However, the higher order increases the variance of min-entropy estimates. We investigate the \emph{bias-variance tradeoff} depending on the order. By considering this tradeoff, we focus on the min-entropy estimator based on the collision entropy (i.e., R{\'{e}}nyi entropy of order two).
	
	\begin{table}[!t]
	\renewcommand{\arraystretch}{1.2}
	\caption{Classification of NIST SP 800-90B Estimators~\cite{Turan2018recommendation,Zhu2020analysis}}
	\vspace{-2mm}
	\label{tab:estimators}
	\centering
	\begin{tabular}{|c|c|}	\hline
		Statistic-based estimator~\cite{Hagerty2012entropy} & Prediction-based estimator~\cite{Kelsey2015predictive} \\ \hline \hline
		Most common value estimator  &  MultiMCW prediction estimator \\ 
		Collision estimator & Lag prediction estimator \\ 
		Markov estimator & MultiMMC prediction estimator \\ 
		Compression estimator & LZ78Y prediction estimator \\ 
		$t$-Tuple estimator &  \\ 
		LRS estimator  & \\ \hline			
	\end{tabular}
	\vspace{-2mm}
	\end{table} 		 
	
	The proposed min-entropy estimator based on the collision entropy is computationally more efficient than other estimators. Furthermore, it has a closed-form solution for the min-entropy estimate, whereas other estimators rely on the bisection method (binary search) to calculate an approximated estimate. Due to its computational efficiency, the min-entropy estimator based on the collision entropy can effectively suppress the variance by including more samples. 
	
	In addition, we propose an \emph{online} estimator that updates the min-entropy estimate as a new sample is received. Note that the compression estimator is inherently an \emph{offline (or batch)} algorithm that requires the whole samples to output the estimate. Unlike the compression estimator, the proposed online estimator can provide a min-entropy estimate with limited samples, then the accuracy of estimates is improved as obtaining more samples. Since the proposed online estimator does not need to store the entire samples, it is proper for applications under stringent resource constraints. 		
	
		 
	The rest of this paper is organized as follows. Section~\ref{sec:prelim} provides an overview of entropies, statistical tests, and the compression estimator. Section~\ref{sec:proposed_Coron} presents a proposed min-entropy estimator based on Coron's test. Section~\ref{sec:proposed_Kim} proposes a min-entropy estimator based on Kim's test and Section~\ref{sec:collision} focuses on the estimator based on the collision entropy. Section~\ref{sec:numerical} provides numerical results and Section~\ref{sec:conclusion} concludes. 
		
	\section{Preliminary: Entropies, Statistical Tests, and Compression Estimator}\label{sec:prelim}	
	
	\subsection{Entropies} \label{sec:entropies}
	
	Suppose that an $N$-bit sample sequence $\vect{s} =(s_1,\ldots,s_N)$ is generated from a given source. The sample sequence $\vect{s}$ is partitioned into non-overlapping $L$-bit blocks as follows: 
	\begin{equation} \label{eq:block}
	\vect{b}(\vect{s}) = \left(b_1, \ldots, b_{\lfloor N/L \rfloor}\right)
	\end{equation}
	where $b_n = (s_{L(n-1)+1}, \ldots, s_{Ln})$ denotes the $n$th block of $\vect{s}$, i.e., $b_n \in \{0, \ldots, B - 1\}$ and $B = 2^L$.  
	
	The Shannon entropy of $L$-bit blocks $\vect{b}(\vect{s})$ is defined as
	\begin{equation} \label{eq:Shannon_entropy}
	H(\mathcal{B}) = H(P) = - \sum_{b = 0}^{B - 1}{p_b \log_2{p_b}}
	\end{equation}
	where $\mathcal{B}$ denotes a random variable over the alphabet $\{0, \ldots, B-1\}$ and $P = (p_0, \ldots, p_{B-1})$ denotes the distribution of $\mathcal{B}$, i.e., $p_b = P(\mathcal{B} = b)$. The corresponding per-bit entropy is given by
	\begin{equation}
	H(\mathcal{S}) = \frac{H(\mathcal{B})}{L}
	\end{equation}  
	where $\mathcal{S}$ denotes the random variable of binary sample $s \in \{0,1\}$. 
	
	The R{\'{e}}nyi entropy of order $\alpha$ is defined as
	\begin{equation} \label{eq:Renyi_entropy}
	H^{(\alpha)}(\mathcal{B}) = H^{(\alpha)}(P) =\frac{1}{1 - \alpha} \log_2{\sum_{b=0}^{B-1}{p_b^{\alpha}}}
	\end{equation} 
	where $\alpha > 0$ and $\alpha \ne 1$. For $\alpha = 2$, the R{\'{e}}nyi entropy corresponds to the \emph{collision} entropy, which is defined as
	\begin{equation} \label{eq:collision_entropy}
	H^{(2)}(\mathcal{B}) = H^{(2)}(P) =-\log_2{\sum_{b=0}^{B-1}{p_b^2}}.
	\end{equation} 
	The corresponding per-bit R{\'{e}}nyi entropy is given by $H^{(\alpha)}(\mathcal{S}) = \frac{H^{(\alpha)}(\mathcal{B})}{L}$. 

	The min-entropy of $\vect{b}(\vect{s})$ is defined as
	\begin{align} 
	H^{(\infty)}(\mathcal{B}) &= H^{(\infty)}(P) = - \log_2{ \left(\max_{b \in \{0, \ldots, B-1\}}p_b \right)} \nonumber \\
	& = -\log_2{\theta} \label{eq:min_entropy}
	\end{align} 
	where $\theta = \max_{b \in \{0, \ldots, B-1\}}p_b$. The corresponding per-bit min-entropy is given by $H^{(\infty)}(\mathcal{S}) =  \frac{H^{(\infty)}(\mathcal{B})}{L}$. 
	
	\begin{remark} \label{rem:entropy_relation}
		The following relations are well known:
		\begin{align}
		H(\mathcal{B}) &= \lim_{\alpha \rightarrow 1} H^{(\alpha)}(\mathcal{B}),  \\
		H^{(\infty)}(\mathcal{B}) &= \lim_{\alpha \rightarrow \infty} H^{(\alpha)}(\mathcal{B}). \label{eq:min_Renyi}
		\end{align}
	\end{remark}

	\begin{remark} \label{rem:Renyi}
		The R{\'{e}}nyi entropy is non-increasing in $\alpha$~\cite{Beck1993thermodynamics}. Hence, $H^{(\infty)}(\mathcal{B}) \le H^{(\alpha)}(\mathcal{B})$, i.e., the min-entropy is the smallest among the R{\'{e}}nyi entropies. 
	\end{remark}

	\subsection{Maurer's Test\label{sec:Maurer}}
	
	Maurer's test is a common randomness test, capable of detecting a wide range of statistical defects~\cite{Maurer1992universal}. It detects whether or not the sequence can be significantly compressed without loss of information~\cite{Rukhin2010statistical,Maurer1992universal}. The formulation of Maurer's test was motivated by the universal source coding algorithms by Elias~\cite{Elias1987interval} and Willems~\cite{Willems1989universal}. Maurer's test is also \emph{universal} since it is designed without knowing the distribution of the source. Maurer's universal test was adopted by the NIST SP 800-22 for the randomness test. The compression estimator of NIST SP 800-90B also relies on Maurer's test.    
	
	Maurer's test takes as input three integers $\{L,Q,K\}$ and an $N$-bit sample $\vect{s}=(s_1, \ldots, s_N)$ where $N = (Q+K) \times L$. The sample sequence $\vect{s}$ is partitioned into non-overlapping $L$-bit blocks $
	\vect{b}(\vect{s}) = (b_1, \ldots, b_{Q+K})$. The first $Q$ blocks are used to initialize the test. The remaining $K$ blocks are used to compute the following test function:
	\begin{equation} \label{eq:Maurer}
	f_\mathcal{M} (\vect{s}) = \frac{1}{K}\sum_{n=Q+1}^{Q+K}{\log_2{D_n(\vect{s})}}
	\end{equation}  
	where $D_n(\vect{s})$ is given by
	\begin{align} 
	&D_n(\vect{s}) \nonumber \\ 
	&= 
	\begin{cases}
	n, & \hspace{-2mm} \text{if } b_{n-i} \ne b_{n}, \forall i <n;\\
	\min\{i: i\ge 1, b_n = b_{n-i} \}, & \hspace{-2mm} \text{otherwise.}
	\end{cases} \label{eq:distance}
	\end{align}
	Note that $D_n(\vect{s})$ is the \emph{minimum distance} between the current $n$th block and any preceding block with the same pattern. Maurer's test can be implemented efficiently as described in~\cite{Maurer1992universal}. 
	
	The size of initial blocks $Q$ should be chosen to be at least $10 \times 2^L$ so as to have a high likelihood that each of the $2^L$ blocks occurs at least once in the first $Q$ blocks. A larger $K$ for test blocks is preferred. It is recommended to use $K \ge 1000 \times 2^L$~\cite{Maurer1992universal,Rukhin2010statistical}. 
	
	Maurer's test is closely related to the Shannon entropy. In~\cite{Maurer1992universal} and \cite{Coron1999accurate}, it was shown that there is a gap between Maurer's test and the Shannon entropy as follows:
	\begin{align} 
	\lim_{L \rightarrow \infty}\left[ \mathbb{E}(f_{\mathcal{M}}(\vect{s})) - H(\mathcal{B}) \right] 
	&\triangleq \int_{0}^{\infty}{e^{-\xi} \log_2 \xi d\xi} \nonumber \\
	&\simeq -0.8327.
	\end{align}
	
	\subsection{Coron's Test for Shannon Entropy\label{sec:Coron}}
		
	Coron's test was proposed to estimate the Shannon entropy by modifying Maurer's test~\cite{Coron1999on}. Coron's test $f_{\mathcal{C}}(\vect{s})$ is given by  
	\begin{equation} \label{eq:Coron}
	f_\mathcal{C} (\vect{s}) = \frac{1}{K}\sum_{n=Q+1}^{Q+K}{g_{\mathcal{C}}(D_n(\vect{s}))}
	\end{equation} 
	where the function $g_{\mathcal{C}}(\cdot)$ should be chosen to satisfy the condition $\mathbb{E}(f_\mathcal{C}(\vect{s})) = H(\mathcal{B})$. Coron showed that the following function $g_{\mathcal{C}}(\cdot)$ achieves this equality condition: 
	\begin{equation}
	g_{\mathcal{C}}(i) = \begin{cases}
	0, & \text{if } i = 1;\\
	\frac{1}{\log{2}} \sum_{k=1}^{i-1}{\frac{1}{k}}, & \text{if } i\ge 2. 
	\end{cases} \end{equation}
	
	The computational complexity of Coron's test is comparable to Maurer's test~\cite{Coron1999on}. To improve the computational efficiency, $g_{\mathcal{C}}(i)$ can be approximated for large $i$ (e.g., $i\ge23$) as follows:
	\begin{equation}
	\sum_{k=1}^{i-1}\frac{1}{k} \simeq \log{(i-1)} + \gamma + \frac{1}{2(i-1)} - \frac{1}{12(i-1)^2}
	\end{equation} 
	where $\gamma$ is Euler's constant, i.e., $\gamma = - \int_{0}^{\infty}{e^{-x} \log{x} dx} \simeq 0.5772$. The detailed algorithm of Coron's test is described in~\cite{Coron1999on}.   
	

	\subsection{Kim's Test for R{\'{e}}nyi Entropy \label{sec:Kim}}
	
	As Coron modified Maurer's test to obtain the Shannon entropy, Kim~\cite{Kim2018low} proposed a variant of Maurer's test to estimate the R{\'{e}}nyi entropy of order $\alpha$. Kim's test $f_{\mathcal{K}}(\vect{s}, \alpha)$ is given by
	\begin{equation} \label{eq:Kim}
	f_\mathcal{K} (\vect{s},\alpha) = \frac{1}{K}\sum_{n=Q+1}^{Q+K}{g_{\mathcal{K}}(D_n(\vect{s}),\alpha)},  
	\end{equation} 
	where $g_{\mathcal{K}}(i, \alpha)$ is defined as
	\begin{equation} \label{eq:g_Kim}
	g_{\mathcal{K}}(i, \alpha) = 
	\begin{cases}
	1, & \text{if } i = 1;\\
	(-1)^{i-1}\cdot \binom{\alpha-2}{i}, & \text{if } i\ge 2. 
	\end{cases}
	\end{equation}
	Here, $\binom{\alpha-2}{i}$ denotes the \emph{generalized} binomial coefficient, i.e., $\binom{\alpha - 2}{i} = \frac{(\alpha-2)_i}{i!}$ where $(\cdot)_i$ is the Poccharmmer symbol. In~\cite{Kim2018low}, it was shown that Kim's test estimates the \emph{power sum} of order $\alpha$ as follows:
	\begin{equation}
	\mathbb{E}(f_\mathcal{K} (\vect{s}, \alpha)) = \sum_{b=0}^{B-1}{p_b^{\alpha}}. 
	\end{equation} 
	Then, the R{\'{e}}nyi entropy of order $\alpha$ can be estimated by
	\begin{equation} \label{eq:collision_Kim}
	H^{(\alpha)}(\mathcal{B}) = \frac{1}{1 - \alpha} \log_2{f_\mathcal{K} (\vect{s}, \alpha)}. 
	\end{equation}
	
	For the \emph{collision} entropy (i.e., R{\'{e}}nyi entropy of $\alpha = 2$), $g_{\mathcal{K}}(i, \alpha = 2)$ is simplified to 
	\begin{equation} \label{eq:g_collision}
	g_{\mathcal{K}}(i, \alpha=2) = 
	\begin{cases}
	1, & \text{if } i = 1;\\
	0, & \text{if } i\ge 2.  
	\end{cases}
	\end{equation}
	Note that $f_\mathcal{K} (\vect{s}, \alpha = 2)$ estimates the collision probability (i.e., $p_c = \sum_{b=0}^{B-1}{p_b^{2}}$) and the collision entropy is given by
	\begin{equation}
	H^{(2)}(\mathcal{B}) = -\log_2{ p_c }. 
	\end{equation}
	
	\begin{remark}\label{rem:collision}
	The collision entropy can be estimated by counting only the case of $D_n(\vect{s}) = 1$ (i.e., the current block $b_n$ is the same as the previous block $b_{n-1}$ which can be interpreted as a \emph{collision} of consecutive samples). Hence, $Q=1$ is sufficient for the initialization of Kim's test, whereas $Q \ge 10 \times 2^L$ are required for the initialization stages of Maurer's test and Coron's test. The computational complexity of Kim's test for the collision entropy is less than those of Maurer's test and Coron's test~\cite[Table 2]{Kim2018low}.  	
	\end{remark}

	\subsection{Compression Estimator of NIST 800-90B}
	
	\begin{algorithm}[!t]
		\caption{The compression estimator of NIST 800-90B~\cite{Turan2018recommendation}} \label{algo:nist}
		\hspace*{\algorithmicindent} \textbf{Input:} $L$-bit blocks $\vect{b}(\vect{s}) = (b_1, \ldots, b_{Q+K})$\\
		\hspace*{\algorithmicindent} \textbf{Output:} $H^{(\infty)}(\mathcal{S})$
		\begin{algorithmic}[1]
			\State Compute $f_\mathcal{M}(\vect{s}) := \frac{1}{K}\sum_{n=Q+1}^{Q+K}{\log_2{D_n(\vect{s})}}$ 
			\State $X := f_\mathcal{M}(\vect{s})$ and $\widehat{\sigma} := c \sqrt{\mathsf{Var}(\log_2 D_n(\vect{s}))}$ 
			\State $X' := X - 2.576 \cdot \frac{\widehat{\sigma}}{\sqrt{K}}$ \label{step:nist_var}  
			\State By the bisection method, solve the following equation for the parameter $\theta \in [\frac{1}{B}, 1]$\label{step:nist_sol}: 
			\begin{equation} \label{eq:key_nist}
				X' = G(\theta) + (B - 1)G(\varphi)
			\end{equation}
			where $G(\cdot)$ is given by \eqref{eq:nist_G} and $\varphi = \frac{1 - \theta}{B - 1}$.			
			\State The estimated per-bit min-entropy is given by
			\begin{equation*}
				H^{(\infty)}(\mathcal{S}) :=
				\begin{cases}
					- \frac{\log_2{\theta}}{L}, & \text{if Step 4 yields a solution}; \\
					1, & \text{otherwise}.
				\end{cases} 						  
			\end{equation*}
		\end{algorithmic}
	\end{algorithm}

	The compression estimator computes Maurer's test and then use it to estimate the lower bound on the min-entropy~\cite{Hagerty2012entropy}. Algorithm~\ref{algo:nist} describes the detailed algorithm of the compression estimator. The NIST SP 800-90B sets $L = 6$ and the corrective factor $c = 0.5907$ in Step 3. Step 3 computes the lower bound of the \unit[99]{\%} confidence interval under the Gaussian assumption.		
	
	\begin{remark}[Corrective Factor]
	The corrective factor $c$ controls the standard deviation of test value since the quantities $D_n(\vect{s})$ are not completely independent. In~\cite{Maurer1992universal}, the corrective factor was determined by a heuristic approximation. Afterward, Coron~\cite{Coron1999accurate} derived an accurate value of corrective factor $c(L, K)$. The NIST SP 800-90B adopts $c = 0.5907$ by setting $L = 6$ and $K \gg 2^L$. For the proposed algorithms (Algorithm~\ref{algo:Coron} and Algorithm~\ref{algo:Kim}), the corrective factors are computed by $c(L, K)$ in~\cite{Coron1999accurate}.  
	\end{remark}
	   	   
	Without loss of generality, we can assume that 
	\begin{equation}\label{eq:p0max}
	p_0 \ge p_1 \ge \ldots \ge p_{B-1}
	\end{equation} 
	where $B = 2^L$. For a given Maurer's test $f_\mathcal{M}(\vect{s})$, the following \emph{near-uniform} distribution can estimate the maximum value of $\theta$, which corresponds to the lower bound on the min-entropy ~\cite{Hagerty2012entropy}: 
	\begin{equation} \label{eq:nearuniform}
	P_{\text{NU}}(b) = 
	\begin{cases}
	\theta, & \text{if } b = 0; \\
	\frac{1 - \theta}{B-1}, & \text{otherwise}.
	\end{cases}
	\end{equation}
	Then, the maximum value of $\theta$ can be obtained from the following equation~\cite{Hagerty2012entropy}:
	\begin{equation} \label{eq:key_nist_original}
	f_{\mathcal{M}}(\vect{s}) =G(\theta) + (B - 1)G(\varphi) 
	\end{equation}	
	where $\varphi = \frac{1- \theta}{B-1}$ and  
	\begin{align} \label{eq:nist_G}
	G(z) &= \frac{1}{K}\sum_{n=Q+1}^{Q+K}{\sum_{i=1}^{n}{F(z,n,i) \cdot \log_2{i}}} \\ 
	F(z, n, i) & = 	
	\begin{cases}
	z^2 (1 - z)^{i-1}, & \text{if } i < n; \\
	z (1-z)^{n-1}, & \text{if } i = n.
	\end{cases} 
	\end{align} 	
	The key equation in Step \ref{step:nist_sol} of Algorithm~\ref{algo:nist} is formulated from \eqref{eq:key_nist_original} by considering the confidence interval. 
	
	
	Algorithm~\ref{algo:nist} should solve the non-closed-form equation by the bisection method. The direct computation of $G(z)$ requires $\mathcal{O}(K^2)$, which can be reduced to $\mathcal{O}(K)$~\cite{SP800_90Bimplementation}. The complexity for solving the key equation is $\mathcal{O}(MK)$ where $M$ corresponds to the number of iterations of the bisection method; $M$ determines the numerical accuracy of the bisection method. The number of samples $K$ would be limited due to the computational complexity. It affects the estimation accuracy because a limited $K$ leads to a large variance $\widehat{\sigma}^2$. We propose a computationally efficient min-entropy estimator so as to include more samples readily and reduce the variance.

	\section{Proposed Estimator Based on Coron's Test}\label{sec:proposed_Coron}
	
	In this section, we propose a min-entropy estimator whose computational complexity is much less than that of the compression estimator while maintaining the estimation accuracy of the compression estimator. 
	

	\subsection{Proposed Estimator Based on Coron's Test}

	We propose a min-entropy estimator by leveraging Coron's test and Fano's inequality. Fano's inequality relates the probability of error and the conditional entropy~\cite{Cover2006}. In~\cite{Han1994generalizing}, Fano's inequality is generalized, and several lower bounds are presented. 
	
	We exploit the relation between the Shannon entropy and the probability of error~\cite{Tebbe1968uncertainty,Golic1987relationship,Feder1994relations}. In the absence of knowledge regarding the random variable $\mathcal{B}$ over the alphabet $\{0, \ldots, B-1\}$, the best guess is the value with the highest probability, i.e., $\theta = \max_{b\in\{0,\ldots,B-1\}}{p_b}$. The minimal error probability $\pi$ in guessing the value of $\mathcal{B}$ is 
	\begin{equation} \label{eq:pi_theta}
		\pi = 1 - \theta. 
	\end{equation} 
	Then, the following inequality is obtained from the Fano's inequality~\cite{Tebbe1968uncertainty,Golic1987relationship,Feder1994relations}:
	\begin{equation} \label{eq:fano}
	H(\mathcal{B}) \le	h(\pi) + \pi \log_2(B-1) 
	\end{equation}
	where  $h(\pi) = -\pi \log_2{\pi} - (1-\pi) \log_2{(1 - \pi)}$. The bound is achieved with equality by the following distribution:
	\begin{equation}
		(p_0, p_1, \ldots, p_{B-1}) = \left(1 - \pi, \frac{\pi}{B-1}, \ldots, \frac{\pi}{B-1}\right),
	\end{equation}
	which is equivalent to the near-uniform distribution of~\eqref{eq:nearuniform}. The bound of \eqref{eq:fano} is \emph{sharp} since the equality is actually achievable~\cite{Tebbe1968uncertainty,Golic1987relationship,Feder1994relations}. Because of $\pi = 1 - \theta$, \eqref{eq:fano} can be modified to 
	\begin{equation}  \label{eq:Fano_modified}
	H(\mathcal{B}) \le h(\theta) + (1 - \theta) \log_2{(B-1)}  
	\end{equation}
	where $h(\pi) = h(\theta)$. 
	
	By using Coron's test and assuming the near-uniform distribution, we can estimate the maximum value of $\theta$ from the following equation:
	\begin{equation}  \label{eq:key_Coron_original}
		f_{\mathcal{C}}(\vect{s}) = h(\theta) + (1 - \theta) \log_2{(B-1)}.   
	\end{equation}
	Note that the maximum value of $\theta$ leads to the lower bound on the actual min-entropy. 
	
	\begin{theorem} \label{thm:Coron_equation} For $\theta \in [\frac{1}{B}, 1]$, there exists only one solution of \eqref{eq:key_Coron_original}. The solution $\theta^*$ decides the sharp lower bound on the min-entropy, i.e., $H^{(\infty)}(\mathcal{B})\ge -\log_2 {\theta^*}$. 		
	\end{theorem}  
	\begin{IEEEproof}
		Suppose that 
		\begin{equation} \label{eq:zeta}
		\zeta(\theta) = h(\theta) + (1-\theta)\log_2{(B-1)}
		\end{equation}
		For $\theta \in (\frac{1}{B}, 1]$, $\zeta(\theta)$ is a strictly decreasing function, i.e., $\zeta(\theta)' = \log_2{\left(\frac{1-\theta}{\theta}\cdot\frac{1}{B-1}\right)} < 0 $. Also, $\zeta(\frac{1}{B})= \log_2{B}$ and $\zeta(1)= 0$. Since $0\le H(\mathcal{B}) \le \log_2{B}$, there exists only one solution $\theta^*$, which is the maximum value that achieves \eqref{eq:Fano_modified} with equality. Hence, $H^{(\infty)}(\mathcal{B}) = -\log_2 {\theta} \ge -\log_2 {\theta^*}$.    
	\end{IEEEproof}	
		 		
	We propose Algorithm~\ref{algo:Coron} by using Coron's test instead of Maurer's test. The key equation of Step \ref{step:Coron_sol} of Algorithm~\ref{algo:Coron} is formulated by~\eqref{eq:key_Coron_original}. The corrective factor of Coron's test is $c= 0.6131$~\cite{Coron1999on}, which is close to $c = 0.5907$ of the compression estimator. Step 3 computes the lower bound of the \unit[99]{\%} confidence interval.  	
		
	\begin{algorithm}[!t]
		\caption{Proposed estimator based on Coron's test} \label{algo:Coron}
		\hspace*{\algorithmicindent} \textbf{Input:} $L$-bit blocks $\vect{b}(\vect{s}) = (b_1, \ldots, b_{Q+K})$\\
		\hspace*{\algorithmicindent} \textbf{Output:} $H^{(\infty)}(\mathcal{S})$
		\begin{algorithmic}[1]
			\State Compute $f_\mathcal{C} (\vect{s}) := \frac{1}{K}\sum_{n=Q+1}^{Q+K}{g_{\mathcal{C}}(D_n(\vect{s}))}$ 
			\State $X := f_\mathcal{C}(\vect{s})$ and $\widehat{\sigma} := c \sqrt{\mathsf{Var}(g_{\mathcal{C}}(D(\vect{s})))}$ 
			\State \label{step:Coron_var}$
			X' := X - 2.576 \cdot \frac{\widehat{\sigma}}{\sqrt{K}}$ 
			\State \label{step:Coron_sol}By the bisection method, solve the following equation for the parameter $\theta \in [\frac{1}{B}, 1]$: 
			\begin{equation} \label{eq:key_Coron}
			X' = h(\theta) + (1 - \theta) \log_2{(B-1)}
			\end{equation}
			\State The estimated per-bit min-entropy is given by
			\begin{equation*}
			H^{(\infty)}(\mathcal{S}) :=
			\begin{cases}
			- \frac{\log_2{\theta}}{L}, & \text{if Step~\ref{step:Coron_sol} yields a solution}; \\
			1, & \text{otherwise}
			\end{cases} 
			\end{equation*}
		\end{algorithmic}
	\end{algorithm}	
	
	\begin{remark}[Computational Efficiency and Estimation Accuracy]
		Unlike the compression estimator, the RHS of \eqref{eq:key_Coron_original} (i.e., $\zeta(\theta)$ of \eqref{eq:zeta}) does not depend on $K$. For a given number of iteration of the bisection method $M$, the complexity of solving \eqref{eq:key_Coron} is $\mathcal{O}(M)$ (see Table~\ref{tab:comparison}). If we store a table for $(\theta, \zeta(\theta))$, then we can readily estimate $\theta^{*} = \arg\min|\zeta(\theta) - f_{\mathcal{C}}(\vect{s})|$. Hence, we can effectively reduce the variance of estimates by including more samples, which improves the estimation accuracy. 
	\end{remark}

	\begin{table*}[!t]
	\renewcommand{\arraystretch}{1.2}
	\caption{Comparison of Compression Estimator and Proposed Estimators}
	\vspace{-2mm}
	\label{tab:comparison}
	\centering
	\begin{tabular}{|c|c|c|c|}	\hline
		& Compression Estimator & Estimator (Coron's Test) & Estimator (Kim's Test) \\ \hline \hline
		Complexity of Test & $\mathcal{O}(K)$ & $\mathcal{O}(K)$ & $\mathcal{O}(K)$ \\\hline
		Complexity of Key Equation  & $\mathcal{O}(MK)$ & $\mathcal{O}(M)$ & $\mathcal{O}(1)$ \\ \hline  						
	\end{tabular}
	\vspace{-2mm}
	\end{table*} 		
	
	The estimated values of the compression estimator and the proposed estimator are almost identical for the same $K$ (see Section~\ref{sec:numerical}). It is mainly because Maurer's test and Coron's test are closely related and both estimates are obtained by assuming the near-uniform distribution. Since the proposed estimator achieves almost identical estimation accuracy with much less computation, the proposed estimator of Algorithm~\ref{algo:Coron} is an appealing alternative to the compression estimator. 
	

	\subsection{Joint Ranges}
	
	Since both the compression estimator (Algorithm~\ref{algo:nist}) and the proposed estimator (Algorithm~\ref{algo:Coron}) estimate the lower bounds on the min-entropy, they are inherently biased estimators. To clarify the underestimate problem, we introduce the \emph{joint range} of a pair of entropies.
		
	For two probability measures $P$ and $Q$, $D_{\phi}(P \| Q)$ denotes the $\phi$-divergence (also known as $f$-divergence)~\cite{Guntuboyina2014sharp,Harremoes2011on}. The joint range of a pair of divergences is defined as follows. 
	
	\begin{definition}[Joint Range of Divergences~\cite{Harremoes2011on}] A point $(x,y) \in [0,\infty]^2$ is $(\phi_1, \phi_2)$-\emph{achievable} if there exist probability measures $P$ and $Q$ such that
		\begin{equation}
			(x,y) = (D_{\phi_1}(P \| Q), D_{\phi_2}(P \| Q)). 
		\end{equation}
	The set of $(\phi_1, \phi_2)$-achievable points is called the \emph{joint range of} $(D_{\phi_1},D_{\phi_2})$. 
	\end{definition}

	Similarly, we define the joint range of a pair of entropies.
	
	\begin{definition}[Joint Range of Entropies] A point $(x,y) \in [0,\infty]^2$ is $(\alpha_1, \alpha_2)$-\emph{achievable} if there exist probability measure $P$ such that
		\begin{equation}
			(x,y) = (H^{(\alpha_1)}(P), H^{(\alpha_2)}(P)). 
		\end{equation}
		The set of $(\alpha_1, \alpha_2)$-achievable points is called the \emph{joint range of} $(H^{(\alpha_1)},H^{(\alpha_2)})$. 
	\end{definition}	 

	Since Coron's test estimates the Shannon entropy, we focus on the joint range of $(H = H^{(1)},H^{(\infty)})$ where $\alpha_1 = 1$ and $\alpha_2 \rightarrow \infty$. In Section~\ref{sec:proposed_Kim}, we investigate the joint range of $(H^{(\alpha)},H^{(\infty)})$ for $\alpha_1 = \alpha > 1$.

%
%

	\subsection{Bias and Gap}

	Suppose that $P$ denotes the true (unknown) distribution of the given sample $\vect{s}$ and $f(\cdot)$ denotes the test function. With a slight abuse of notation, we set $x = f(\vect{s}) = f(P)$, i.e., $x$ represents the test value. Then, the \emph{bias} is defined as
	\begin{equation} \label{eq:bias}
		\mathsf{bias}(P,f,x) = H^{(\infty)}(P) - \min_{Q: f(Q) = x} H^{(\infty)}(Q).
	\end{equation}
	For Maurer's test and Coron's test, the lower bounds are achieved by the near-uniform distribution $P_{\text{NU}}$, i.e., 
	\begin{align} \label{eq:bias2}
		\mathsf{bias}(P,f,x) &= H^{(\infty)}(P) - H^{(\infty)}(P_{\text{NU}}) 
	\end{align}
	where $x = f(P_{\text{NU}})$. 

	Since the true distribution $P$ is unknown, we replace $H^{(\infty)}(P)$ by its maximum and define the \emph{gap} as follows: 
	\begin{equation} \label{eq:gap}
		\mathsf{gap}(f, x) = \max_{Q: f(Q) = x} H^{(\infty)}(Q) - \min_{Q: f(Q) = x} H^{(\infty)}(Q),
	\end{equation}
	which represents the \emph{maximum bias}. 

	For a given Maurer's test value, it was shown that $\max_{Q: f(Q) = x} H^{(\infty)}(Q)$ is achieved by the following \emph{inverted} near-uniform distribution~\cite{Hagerty2012entropy}:  
	\begin{equation} \label{eq:inverted_nearuniform}
		P_{\text{INU}}(b) = 
		\begin{cases}
			\psi, & \text{if}~b \in \left\{0, \ldots, \left\lfloor \frac{1}{\psi} \right\rfloor - 1 \right\}; \\
			1 - \left\lfloor \frac{1}{\psi} \right\rfloor \psi, & \text{if}~b = \left\lfloor \frac{1}{\psi} \right\rfloor; \\
			0,& \text{otherwise}
		\end{cases}
	\end{equation}
	where $\psi = \max_{b\in\{0,\ldots,B-1\}}{P_{\text{INU}}(b)}$. 

	Since Coron's test estimates the Shannon entropy, we claim that $P_{\text{INU}}$ achieves $\max_{Q:f(Q)=x} H^{(\infty)}(Q)$ based on~\cite{Tebbe1968uncertainty,Feder1994relations}. Hence, the gaps for both Maurer's test and Coron's test become
	\begin{align} \label{eq:gap2}
		\mathsf{gap}(f, x) &=  H^{(\infty)}(P_{\text{INU}}) - H^{(\infty)}(P_{\text{NU}}) 
	\end{align}
	where $x = f(P_{\text{INU}})=f(P_{\text{NU}})$.  

\begin{figure}[!t]
		\centering
		\subfloat[]{\includegraphics[width=0.4\textwidth]{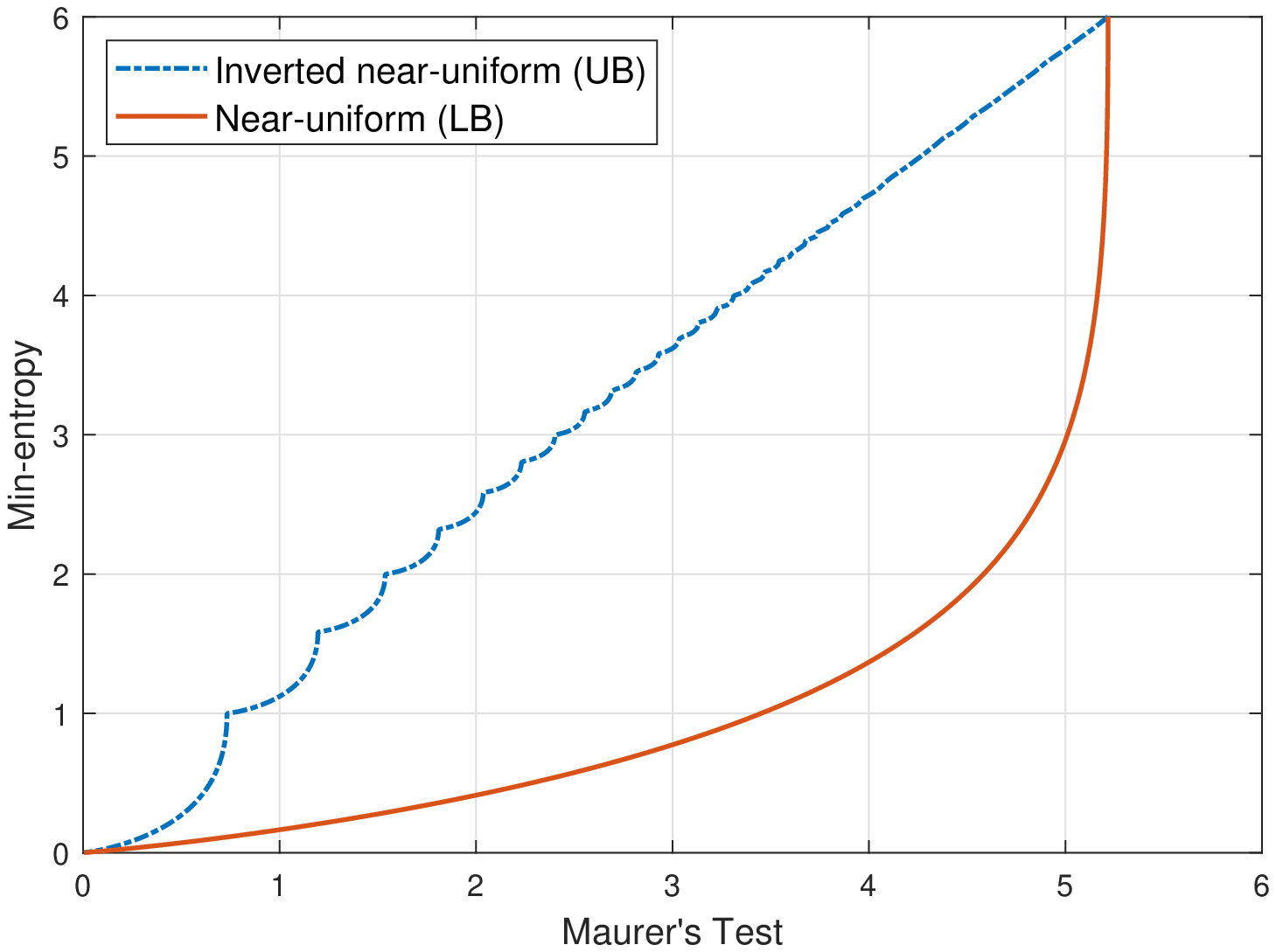}
			\label{fig:gap_a}}
		\hfil
		\subfloat[]{\includegraphics[width=0.4\textwidth]{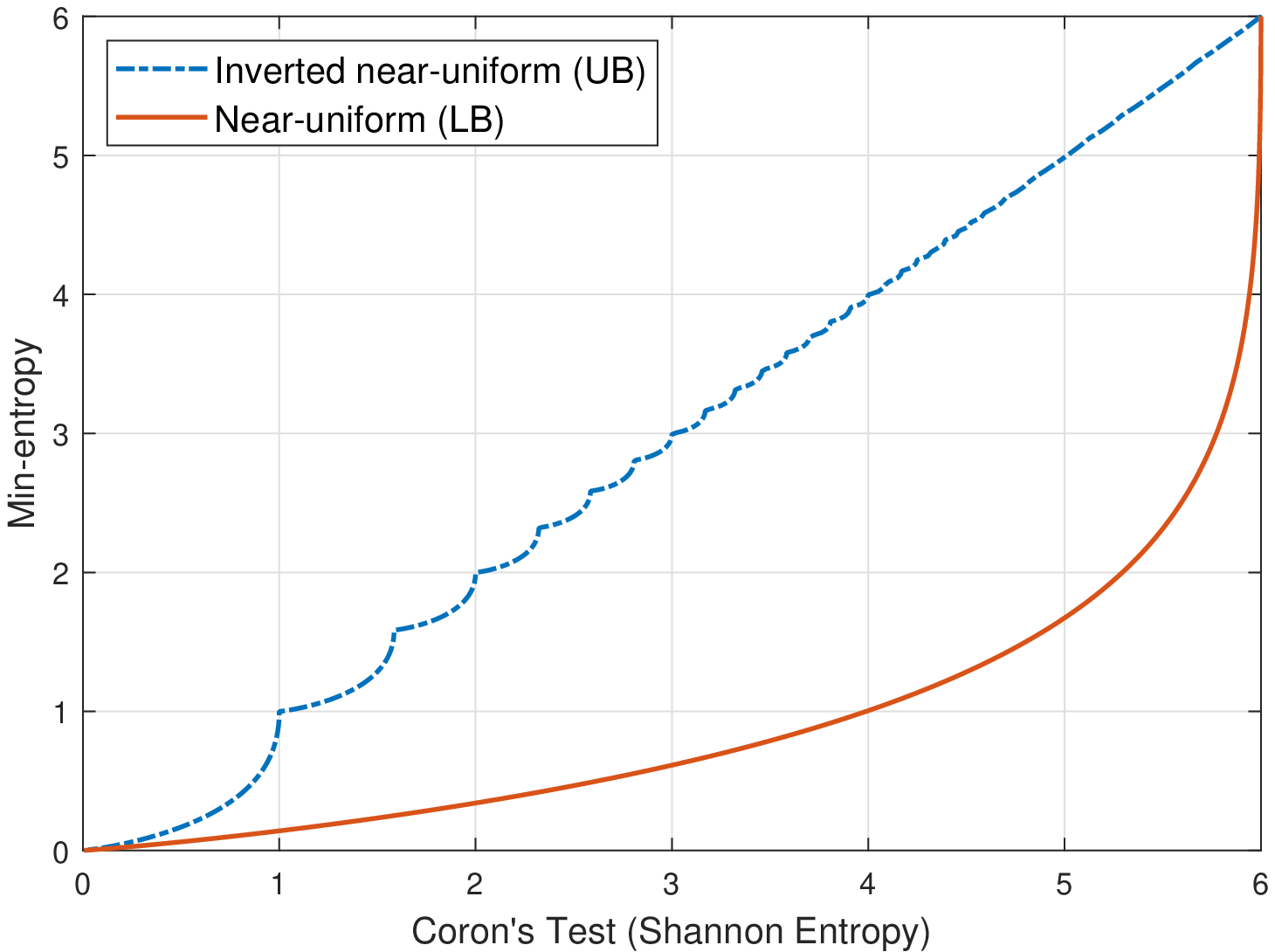}	
			\label{fig:gap_b}}
		\caption{The joint range of test values and min-entropies for $L=6$: (a) Compression estimator based on Maurer's test and (b) proposed estimator based on Coron's test. The maximum value of Maurer's test is 5.2177~\cite{Maurer1992universal} and the maximum value of Coron's test is $L$ (i.e., the maximum value of Shannon entropy).}
		\label{fig:gap}
	\end{figure}

	Fig.~\ref{fig:gap} shows the joint ranges corresponding to the compression estimator and the proposed estimator. In particular, Fig.~\ref{fig:gap}\subref{fig:gap_b} corresponds to the joint range of $(H(P), H^{(\infty)}(P))$ where $\alpha_1 = 1$ and $\alpha_2 \rightarrow \infty$. We note that the gaps cannot be tightened because the lower and upper bounds are achieved by the near-uniform distribution and the inverted near-uniform distribution, respectively. 
	
	\begin{remark} \label{rem:extreme}
	The gap and bias become zero for only two extreme points, i.e., $H^{(\infty)}(\mathcal{B})=0$ and $H^{(\infty)}(\mathcal{B})=L$. 
	\end{remark}

	Since most sample sequences would not correspond to these two extreme points, the estimated min-entropies by the compression estimator and the proposed estimator can be significant underestimates. 		
	
	\section{Proposed Estimator Based on Kim's Test}\label{sec:proposed_Kim}
	
	In this section, we propose a min-entropy estimator based on the R{\'{e}}nyi entropy. We show that the proposed estimator can effectively reduce the bias and be computationally efficient.  	
	
	\subsection{Proposed Estimator Based on Kim's Test}	
		
	For a given R{\'{e}}nyi entropy, we can estimate the lower bound on the min-entropy by assuming the near-uniform distribution as in the compression estimator and the proposed estimator based on Coron's test. 
	
	\begin{lemma}[{\cite[Theorem 6]{Ben-Bassat2978renyi}}]\label{lem:Fano_Renyi} Suppose that $\theta = \max_{b\in\{0,\ldots,B-1\}}{p_b}$. Then, the following inequality holds:   
		\begin{equation} \label{eq:Fano_Renyi}
		H^{(\alpha)}(\mathcal{B}) \le \frac{1}{1 - \alpha} \log_2{\left(\theta^{\alpha} +  \frac{(1-\theta)^\alpha}{(B-1)^{\alpha-1}}\right)} 
		\end{equation} 
		for $\alpha \ne 1$. The near-uniform distribution of \eqref{eq:nearuniform} achieves this bound with equality. 
	\end{lemma}



	By assuming the near-uniform distribution as in \eqref{eq:key_Coron_original}, we can estimate the maximum value of $\theta$ from the following equation: 
	\begin{equation}\label{eq:key_Kim_original}
		f_{\mathcal{K}}(\vect{s},\alpha) = \theta^{\alpha} +  \frac{(1-\theta)^\alpha}{(B-1)^{\alpha-1}} 
	\end{equation}
	where $2^{(1 - \alpha)H^{(\alpha)}(\mathcal{B})} = f_{\mathcal{K}}(\vect{s},\alpha)$ because of \eqref{eq:collision_Kim}. The following theorem shows that the lower bound on the min-entropy can be estimated by Lemma~\ref{lem:Fano_Renyi} and Kim's test.  
	
	\begin{theorem} \label{thm:Kim_equation} For $\theta \in [\frac{1}{B}, 1]$ and $\alpha > 1$, there exists only one solution of \eqref{eq:key_Kim_original}. The solution $\theta^*$ decides the sharp lower bound on the min-entropy, i.e., $H^{(\infty)}(\mathcal{B})\ge -\log_2 {\theta^*}$. 		
	\end{theorem}  
	\begin{IEEEproof}
		Suppose that $\zeta(\theta)=\theta^{\alpha} + \frac{(1 - \theta)^{\alpha}}{(B-1)^{\alpha - 1}}$. For $\theta \in (\frac{1}{B}, 1]$, $\zeta(\theta)$ is a strictly increasing function, i.e., $\zeta(\theta)' > 0$. Also, $\zeta(\frac{1}{B})= B^{1 - \alpha}$ and $\zeta(1)= 1$. Since $0\le  H^{(\alpha)}(\mathcal{B}) \le \log_2{B}$, we observe that $B^{1 - \alpha} \le f_{\mathcal{K}}(\vect{s},\alpha)  \le 1$. Hence, there exists only one solution $\theta^*$, which is the maximum value that achieves \eqref{eq:Fano_Renyi} with equality. Hence, $H^{(\alpha)}(\mathcal{B}) = -\log_2 {\theta} \ge -\log_2 {\theta^*}$.    
	\end{IEEEproof}


	Based on Theorem~\ref{thm:Kim_equation}, we propose Algorithm~\ref{algo:Kim} to estimate the min-entropy. As in Algorithm~\ref{algo:nist} and Algorithm~\ref{algo:Coron}, the key equation of Step~\ref{step:Kim_key} is formulated from \eqref{eq:key_Kim_original} by considering the confidence level of \unit[99]{\%}. The corrective factor $c$ of Kim's test depends on $\alpha$ as well as $L$ and $K$, which can be computed as in~\cite{Coron1999accurate}.  

	\begin{algorithm}[!t]
		\caption{Proposed estimator based on Kim's test} \label{algo:Kim}
		\hspace*{\algorithmicindent} \textbf{Input:} $L$-bit blocks $\vect{b}(\vect{s}) = (b_1, \ldots, b_{Q+K})$\\
		\hspace*{\algorithmicindent} \textbf{Output:} $H^{(\infty)}(\mathcal{S})$
		\begin{algorithmic}[1]
			\State Compute $\label{step:Kim}
			f_\mathcal{K} (\vect{s},\alpha) := \frac{1}{K}\sum_{n=Q+1}^{Q+K}{g_{\mathcal{K}}(D_n(\vect{s}),\alpha)}$ 
			\State $X := f_\mathcal{K}(\vect{s},\alpha)$ and $\widehat{\sigma} := c \sqrt{\mathsf{Var}(g_{\mathcal{K}}(D(\vect{s}),\alpha))}$ 
			\State $X' := X - 2.576 \cdot \frac{\widehat{\sigma}}{\sqrt{K}}$ 
			\State \label{step:Kim_key}By the bisection method, solve the following equation for the parameter $\theta \in [\frac{1}{B}, 1]$: 
			\begin{equation} \label{eq:key_Kim}
				X' = \theta^{\alpha} +  \frac{(1-\theta)^\alpha}{(B-1)^{\alpha-1}} 
			\end{equation}
			\State The estimated per-bit min-entropy is given by
			\begin{equation*}
			H^{(\infty)}(\mathcal{S}) :=
			\begin{cases}
			- \frac{\log_2{\theta}}{L}, & \text{if Step~\ref{step:Kim_key} yields a solution}; \\
			1, & \text{otherwise}
			\end{cases} 
			\end{equation*}
		\end{algorithmic}
	\end{algorithm}	

	\subsection{Bias-Variance Tradeoff} \label{sec:bias_var}
	
	We investigate the \emph{bias-variance tradeoff} depending on the order of R{\'{e}}nyi entropy. A higher order of R{\'{e}}nyi entropy reduces the bias and increases the variance of estimates. Conversely, a lower order reduces the variance and increases the bias. 
	
\begin{figure}[!t]
	\centering
	\includegraphics[width=0.4\textwidth]{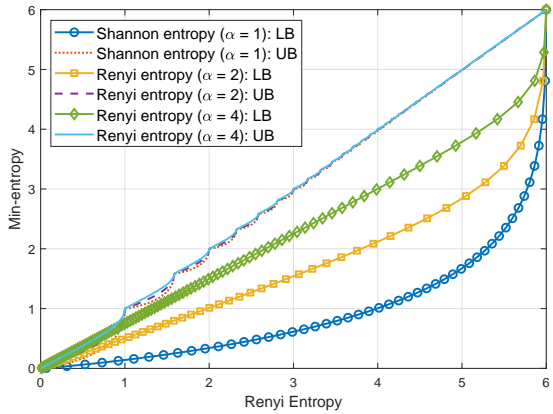}
	\caption{The joint range of $(H^{(\alpha)}(P), H^{(\infty)}(P))$ for $L = 6$.}
	\label{fig:gap_order}
	\end{figure}

	Fig.~\ref{fig:gap_order} shows the joint range of $(H^{(\alpha)}(P), H^{(\infty)}(P))$.	We observe that the gap (i.e., the maximum bias) can be suppressed by increasing the order of R{\'{e}}nyi entropy. For a given R{\'{e}}nyi entropy, the lower and uppers bounds on the min-entropy are achieved by the near-uniform distribution and the inverted near-uniform distribution, respectively. The gap decreases for the higher order $\alpha$. It is because the lower bound can be significantly increased by the higher order, whereas the changes of upper bounds are limited. Eventually, the gap converges to 0 as $\alpha \rightarrow \infty$.

	We show that the bias of min-entropy estimation can be reduced by increasing the order of R{\'{e}}nyi entropy. 	
	
	\begin{theorem} \label{thm:order_gap}
		Suppose that $\theta^{(\alpha)}$ and $\theta^{(\alpha+1)}$ are estimated values in Algorithm~\ref{algo:Kim} by using $f_{\mathcal{K}}(\vect{s},\alpha)$ and $f_{\mathcal{K}}(\vect{s},\alpha+1)$, respectively. If $\theta^{(\alpha)} \gg \frac{1}{1 + (B-1)^{\frac{\alpha - 1}{\alpha}}}$ and $\alpha > 1$, then     
		\begin{equation} \label{eq:order_improve}
		H^{(\infty)}(P) \ge \log_2{\frac{1}{\theta^{(\alpha + 1)}}} \ge  \log_2{\frac{1}{\theta^{(\alpha)}}}. 
		\end{equation}
		Hence, the bias can be reduced by increasing the order of R{\'{e}}nyi entropy.
	\end{theorem}
	\begin{IEEEproof}
	The proof is given in Appendix~\ref{pf:order_gap}. 		  
	\end{IEEEproof}

	\begin{figure}[t]
	\centering
	\includegraphics[width=0.4\textwidth]{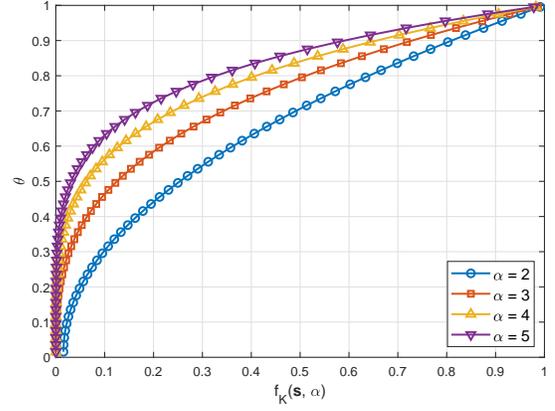}
	\caption{The relation between Kim's test $f_{\mathcal{K}}(\vect{s},\alpha)$ and the estimated $\theta$ of \eqref{eq:key_Kim_original} for $L=6$.} 
	\label{fig:var}
	\end{figure}	

	\begin{figure}[t]
	\centering
	\includegraphics[width=0.4\textwidth]{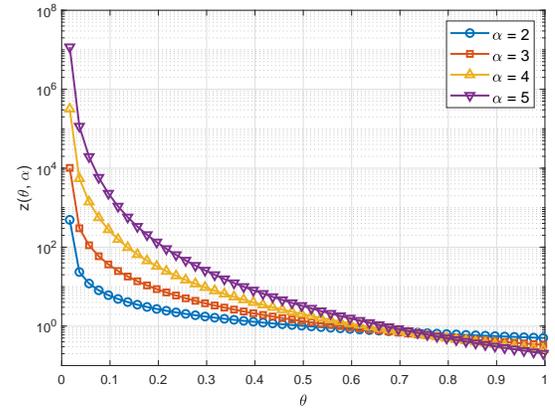}
	\caption{The relation between $\theta$ and $z(\theta, \alpha)$ of \eqref{eq:def_z} for $L=6$ and $\theta \in \left[\frac{1}{B} + \delta, 1\right]$ where $\delta = 0.001$.}
	\label{fig:var_slope}
	\end{figure}	

	Now, we investigate how the order affects the variance of estimates. Since we estimate $\theta$ from $f_{\mathcal{K}}(\vect{s}, \alpha)$ in Algorithm~\ref{algo:Kim}, we consider the relation between $f_{\mathcal{K}}(\vect{s}, \alpha)$ and $\theta$ as shown in Fig.~\ref{fig:var}. Then, 
	\begin{align} 
		\mathsf{Var}(\theta^{(\alpha)}) &= z(\theta,\alpha)^2 \cdot \mathsf{Var}(f_{\mathcal{K}}(\vect{s}, \alpha)) \\
		& = z(\theta,\alpha)^2 \cdot \frac{c^2}{K} \cdot \mathsf{Var}(g_{\mathcal{K}}(D, \alpha)) \label{eq:var_2}
	\end{align}	
	where $D$ denotes the random variable of $D_n(\vect{s})$ in \eqref{eq:Kim} and $c$ denotes the corrective factor. Also, $z(\theta, \alpha)$ denotes the derivative $\frac{d\theta}{d f_{\mathcal{K}}(\vect{s}, \alpha)}$ as follows: 
	\begin{equation} \label{eq:def_z}
		z(\theta, \alpha) = \frac{d\theta}{d f_{\mathcal{K}}(\vect{s}, \alpha)} = \frac{1}{\alpha \left\{ \theta^{\alpha - 1} -  \left(\frac{1 - \theta}{B-1} \right)^{\alpha - 1} \right\}}.  
	\end{equation}	
	Note that $\mathsf{Var}(\theta^{(\alpha)}) \rightarrow \infty$ as $\theta \rightarrow \frac{1}{B}$. It is because $z(\theta, \alpha) \rightarrow \infty$ as $\theta \rightarrow \frac{1}{B}$ by \eqref{eq:def_z}. 
	
	We observe that $z(\theta, \alpha)$ dominates the change of $\mathsf{Var}(\theta^{(\alpha)})$. The reason is that $z(\theta, \alpha)$ increases significantly with $\alpha$ for high, or even moderate values of min-entropy as shown in Fig.~\ref{fig:var_slope}. The following theorem shows that $z(\theta, \alpha)$ can increase exponentially with $\alpha$. 

	\begin{theorem}\label{thm:slope}
		For $\theta = \frac{1}{B} + \delta$ where $\delta \ll \frac{1}{B}$, $z(\theta, \alpha)$ is approximated to
		\begin{equation}
		z(\theta, \alpha) \simeq \frac{B^{\alpha-3}}{\alpha (\alpha - 1)} \cdot \frac{B-1}{\delta} 
		\end{equation}
		and
		\begin{equation} \label{eq:xi}
		\xi = \frac{z(\theta, \alpha + 1)}{z(\theta, \alpha)} \simeq \frac{\alpha-1}{\alpha + 1}\cdot B.   
		\end{equation}
	For $B = 64$ (i.e., the given parameter of NIST SP 800-90B), $\xi > 1$ (i.e., $z(\theta, \alpha) < z(\theta, \alpha + 1)$) if $\alpha  > \frac{65}{63}$. 	
	\end{theorem}
	\begin{IEEEproof}
		The proof is given in Appendix~\ref{pf:slope}. 
	\end{IEEEproof}

	Next, we investigate the relation between $\alpha$ and $\mathsf{Var}(g_{\mathcal{K}}(D, \alpha))$. 
	\begin{lemma}\label{lem:var_g}
		For a given sample sequence $\vect{s}$, 
		\begin{equation}
		\mathsf{Var}(g_{\mathcal{K}}(D, \alpha=2)) \le \mathsf{Var}(g_{\mathcal{K}}(D, \alpha=3)).
		\end{equation}
	\end{lemma}
	\begin{IEEEproof}
		The proof is given in Appendix~\ref{pf:var_g}. 		
	\end{IEEEproof}			

	Even for $\alpha \ge 3$, we can compute $\mathsf{Var}(g_{\mathcal{K}}(D, \alpha))$ by using \eqref{eq:g_Kim}. Table~\ref{tab:var_c} shows numerical values of $\mathsf{Var}(g_{\mathcal{K}}(D, \alpha))$ and the corrective factor $c$ for the uniform distribution. Although the changes of $\mathsf{Var}(g_{\mathcal{K}}(D, \alpha))$ and $c$ are much less than the change of $z(\theta, \alpha)$, the higher order leads to the larger values of $c^2 \cdot \mathsf{Var}(g_{\mathcal{K}}(D, \alpha))$ in \eqref{eq:var_2}. 
	
	\begin{table}[!t]
		\renewcommand{\arraystretch}{1.2}
		\caption{$\mathsf{Var}(g_{\mathcal{K}}(D, \alpha))$ and Corrective Factor for Uniform Distribution}
		\vspace{-2mm}
		\label{tab:var_c}
		\centering
		\begin{tabular}{|c|c|c|}	\hline
			$\alpha$ & $\mathsf{Var}(g_{\mathcal{K}}(D, \alpha))$ & Corrective factor $c$ \\ \hline \hline
			2 & 0.0154 & 1 \\\hline
			3 & 0.0310 & 1.008 \\\hline			
			4 & 0.0923 & 1.008 \\\hline				
			5 & 0.3052 & 1.008 \\\hline					
		\end{tabular}
		\vspace{-2mm}
	\end{table} 	
		
	Finally, we show that $\mathsf{Var}(\theta^{(2)}) < \mathsf{Var}(\theta^{(3)})$ for most sample sequences. Note that $\mathsf{Var}(\theta^{(2)}) < \mathsf{Var}(\theta^{(3)})$ is equivalent to $\sigma^{(2)} < \sigma^{(3)}$ where $\sigma^{(\alpha)}$ denotes the standard deviation of $H^{(\infty)}(\mathcal{B})$ estimated by $f_{\mathcal{K}}(\vect{s}, \alpha)$.
	
	\begin{theorem} \label{thm:var} For a sample sequence $\vect{s}$ with $\theta$, $\mathsf{Var}(\theta^{(2)}) < \mathsf{Var}(\theta^{(3)})$ if
		\begin{equation} \label{eq:condition_theta_var}
			\theta < \frac{2}{3} - \frac{1}{3(B-2)}.
		\end{equation}
	\end{theorem}  		
	\begin{IEEEproof}The proof is given in Appendix~\ref{pf:var}. 
	\end{IEEEproof}
			
	\begin{remark}
		For $B = 64$ (i.e., the given parameter of NIST SP 800-90B), this condition of $\theta$ corresponds to $\theta < \frac{123}{186} \simeq 0.6613$ and $H^{(\infty)}(\mathcal{S}) > 0.0994$. Hence, we claim that $\sigma^{(2)} < \sigma^{(3)}$ for the most sample sequences.  
	\end{remark}
	
%

	Due to the bias-variance tradeoff, we observe that $\alpha = 2$ is a proper value from numerical evaluations in Section~\ref{sec:numerical}. In the following section, we propose estimation algorithms based on Kim's test for the collision entropy. 

	\section{Proposed Estimator Based on Collision Entropy} \label{sec:collision}

	\subsection{Proposed Estimator Based on Collision Entropy}

	For the collision entropy, we show that the proposed estimator of Algorithm~\ref{algo:Kim} has the following advantages: 
	\begin{enumerate}
		\item The computations are simplified because a closed-form solution of Step~\ref{step:Kim_key} can be derived (see Corollary~\ref{cor:closed_sol}); 
		\item Samples for initialization are not required. Note that both the compression estimator and the proposed estimator based on Coron's test require $Q (> 10 \times 2^L)$ samples for initialization (see Remark~\ref{rem:collision}).
	\end{enumerate}

	\begin{corollary}\label{cor:closed_sol}
		For a estimated collision entropy $H^{(2)}(\mathcal{B}) = f_{\mathcal{K}}(\vect{s},\alpha=2)$, the min-entropy is lower bounded as follows:
		\begin{equation}
		H^{(\infty)}(\mathcal{B}) \ge -\log_2{\theta^{(2)}}
		\end{equation}
		where 
		\begin{align} \label{eq:collision_theta}
		&\theta^{(2)} = \nonumber \\
		&\begin{cases}
		\frac{1}{B}, & \text{if } 0 \le f_{\mathcal{K}}(\vect{s},2) \le \frac{1}{B}; \\
		\frac{1 + \sqrt{(B-1)(B \cdot f_{\mathcal{K}}(\vect{s},2) - 1)}}{B}, & \text{if } \frac{1}{B} < f_{\mathcal{K}}(\vect{s},2) \le 1
		\end{cases} 
		\end{align}
	where $f_{\mathcal{K}}(\vect{s},2) = f_{\mathcal{K}}(\vect{s},\alpha=2)$.  
	\end{corollary}
	\begin{IEEEproof} First, we note that $0 \le f_{\mathcal{K}}(\vect{s},2) \le 1$ by \eqref{eq:Kim} and \eqref{eq:g_collision}.
	If $f_{\mathcal{K}}(\vect{s},2) \le \frac{1}{B}$, then we set $f_{\mathcal{K}}(\vect{s},2) = \frac{1}{B}$ because $H^{(2)}(\mathcal{B}) = -\log_2{f_{\mathcal{K}}(\vect{s},2)} \le L$ by entropy definition. Hence, $\theta^{(2)} = \frac{1}{B}$. If $\frac{1}{B} < f_{\mathcal{K}}(\vect{s},2) \le 1$, then we derive $\theta^{(2)} = \frac{1 \pm \sqrt{(B-1)(B \cdot f_{\mathcal{K}}(\vect{s},2) - 1)} }{B}$ from \eqref{eq:key_Kim_original}. We choose $\theta^{(2)} = \frac{1 + \sqrt{(B-1)(B \cdot f_{\mathcal{K}}(\vect{s},2) - 1)} }{B}$ because of the given condition of $\frac{1}{B} \le \theta^{(2)} \le 1$.    
	\end{IEEEproof}	

	It is worth mentioning that the proposed estimator based on the collision entropy has advantages over the compression estimator in terms of bias (tightened gap), computational complexity (closed-form solution), and data efficiency (skipped initialization stage) at the cost of variance.  
	
	\subsection{Online Estimator Based on Collision Entropy}
	
	We propose an \emph{online} estimator by leveraging the advantages of the collision entropy (see Algorithm~\ref{algo:online}). Since the proposed online estimator processes samples in an online manner, it can estimate the min-entropy with limited samples and then improve its estimation accuracy as getting more samples. Since the proposed online estimator does not need to store the entire samples, it is lightweight and proper for applications with stringent resource constraints. 		
	
	The proposed online estimator has two parts: 1) Estimation of the collision probability $p_c$; 2) estimation of the min-entropy from the collision probability. The first part (Steps~\ref{step:collision_start}--\ref{step:collision_end}) is an online algorithm to estimate the collision probability. For the collision probability, $g_{\mathcal{K}}(i,2)$ is given by \eqref{eq:g_collision}. Hence, it counts only an event that a new block is the same as its previous one (Step~\ref{step:collision_count}) (i.e., collision counting in consecutive blocks). Then, the collision probability $p_c$ converges to $f_{\mathcal{K}}(\vect{s},2)$ as getting more blocks.
	
	The second part (Steps~\ref{step:min_start}--\ref{step:min_end}) estimates the min-entropy from the collision probability $p_c$ in an online manner. This part relies on the closed-form solution of $\theta$ in Corollary~\ref{cor:closed_sol}. The proposed online algorithm is computationally simple and can output a new estimate of min-entropy $H^{(\infty)}_k(\mathcal{S})$ as getting a new block $b_k$.     
	
	\begin{algorithm} [t]
		\caption{Proposed \emph{online} min-entropy estimator} \label{algo:online}
		\hspace*{\algorithmicindent} \textbf{Input:} $L$-bit blocks $\vect{b}(\vect{s}) = (b_1, \ldots, b_K)$\\
		\hspace*{\algorithmicindent} \textbf{Output:} 	$H^{(\infty)}_k(\mathcal{S})$ for $k \in \{1,\ldots,K\}$ and the collision index set $\mathcal{C}$	
		\begin{algorithmic}[1]
			\State $k := 1$, $c := 0$, $v: = b_1$, $\mathcal{C} = \emptyset$ \Comment{Initialization}
			\While {$k < K$} 
			\State $k: = k+1$ \label{step:collision_start}
			\State $u: = b_k$ 
			\If {$u = v$}
				\State $c: = c + 1$  \Comment{Count collision} \label{step:collision_count}
				\State $\mathcal{C}:= \mathcal{C} \cup k$
			\EndIf
			\State $v : = u$ 
			\State $p_c := \frac{c}{k}$ \Comment{Compute collision probability} \label{step:collision_end}
			\If {$p_c > \frac{1}{B} $} \label{step:min_start}
				\State $\theta: = \frac{1 + \sqrt{(B-1)(B \cdot p_c -1})}{B}$
			\ElsIf{$p_c \le \frac{1}{B} $}
				\State $\theta: = \frac{1}{B}$
			\EndIf
			\State $H^{(\infty)}_k(\mathcal{S}) := -\frac{\log_2{\theta}}{L}$ \label{step:min_end}
			\EndWhile
		\end{algorithmic}
	\end{algorithm}	

	The proposed online estimator is helpful to detect low entropy sources with limited samples. It is because the estimation variance of low entropy sources is not large, so its estimate can be obtained reliably with limited samples (see Fig.~\ref{fig:online}). Hence, the proposed online estimator can filter out the low entropy sources very effectively.

	\section{Numerical Results}\label{sec:numerical}
	
	We evaluate our proposed estimators for simulated and real-world data samples. We compare the proposed estimators to the compression estimator since they rely on the same statistics (i.e., minimum distances between the matching blocks $D_n(\vect{s})$ of \eqref{eq:distance}). 
	
	Datasets of simulated samples are produced using the following distribution families as in~\cite{Kelsey2015predictive}:
	\begin{itemize}
		\item \emph{Binary memoryless source (BMS):} Samples are generated by Bernoulli distribution with $P(\mathcal{S}=1) = p$ and $P(\mathcal{S}=0) = 1 - p$ (IID);   
		\item \emph{Near-uniform distribution:} Samples are generated by near-uniform distribution of \eqref{eq:nearuniform} (IID); 
		\item \emph{Inverted near-uniform distribution:} Samples are generated by inverted near-uniform distribution of \eqref{eq:inverted_nearuniform} (IID); 
		\item \emph{Normal distribution rounded to integers:} Samples are drawn from a normal distribution and rounded to integer values (IID); 
		\item \emph{Markov model:} Samples are generated using the first order Markov model (non-IID).  
	\end{itemize}

	One hundred simulated sources were created in each of the above datasets. A sequence of 6,000,000 bits (1,000,000 blocks) was generated from each simulated source. Note that the compression estimator of NIST SP 800-90B sets $L=6$. For each source, the correct min-entropy is derived from the given probability distribution as in~\cite{Kelsey2015predictive}.   


	\begin{table*}[t]
	\renewcommand{\arraystretch}{1.2}
	\caption{Error Measures for BMS with $p$}
	\vspace{-2mm}
	\label{tab:bms}
	\centering
	\begin{tabular}{|c|c|c|c|c|c|}	\hline
		& $p=0.1$ & $p=0.2$ & $p=0.3$ & $p=0.4$ & $p=0.5$ \\ \hline \hline
		MSE of Algo.~\ref{algo:nist}  & 0.0043 & 0.0179 & 0.0365 & 0.0434 & 0.0105 \\ \hline
		MSE of Algo.~\ref{algo:Coron} & 0.0036 & 0.0157 & 0.0336 & 0.0416 & 0.0107 \\ \hline
		MSE of Algo.~\ref{algo:Kim} ($\alpha=2$) & 0.0001 & 0.0012 & 0.0068 & 0.0184 & 0.0217 \\ \hline
		\hline
		MPE of Algo.~\ref{algo:nist}  & 43.09 & 41.56 & 37.14 & 28.26 & 10.10 \\ \hline
		MPE of Algo.~\ref{algo:Coron} & 39.23 & 38.96 & 35.65 & 27.67 & 10.21 \\ \hline
		MPE of Algo.~\ref{algo:Kim} ($\alpha=2$) & 5.29 & 10.86 & 16.05 & 18.42 & 14.62 \\ \hline
	\end{tabular}
	\vspace{-2mm}
\end{table*} 		
		
	\begin{figure}[t]
	\centering
	\includegraphics[width=0.4\textwidth]{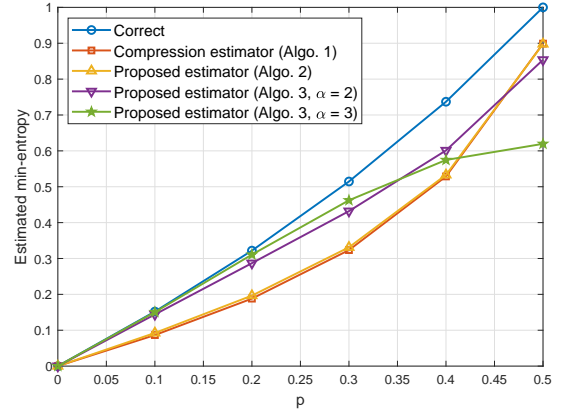}
	\caption{Comparison of min-entropy estimators for binary memoryless sources with $p$.}
	\label{fig:bms}
	\vspace{-4mm}
	\end{figure}

	\begin{figure}[t]
	\centering
	\subfloat[]{\includegraphics[width=0.4\textwidth]{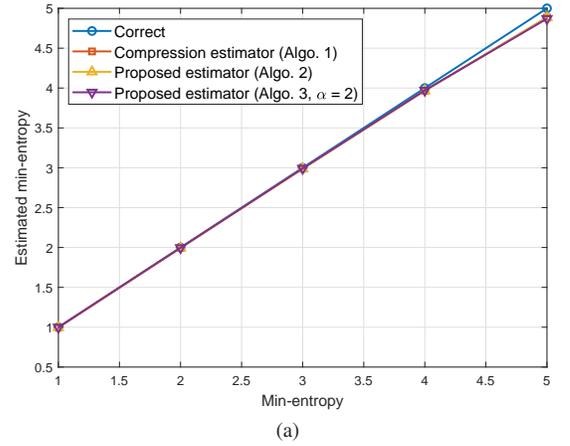}
		\label{fig:near}}
	\vspace{-2mm}
	\hfil
	\subfloat[]{\includegraphics[width=0.4\textwidth]{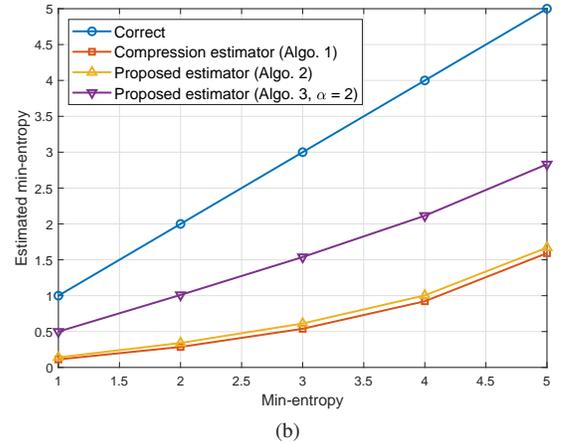}
		\vspace{-4mm}
		\label{fig:inv}}
	\caption{Comparison of min-entropy estimators for (a) near-uniform distributed sources and (b) inverted near-uniform distributed sources.}
	\label{fig:near_inv}
\end{figure}		

	Fig.~\ref{fig:bms} compares the min-entropy estimators for BMS with $p$. The correct min-entropy is given by $H^{(\infty)}(\mathcal{S}) = -\log_2{\max\{p, 1-p\}}$. We observe that the compression estimator (Algorithm~\ref{algo:nist}) of NIST SP 800-90B and the proposed estimator based on Coron's test (Algorithm~\ref{algo:Coron}) are almost identical. By comparing the computational complexities of estimators (see Table~\ref{tab:comparison}), the proposed estimator based on Coron's test is an appealing alternative to the compression estimator. 
	
	The proposed estimators based on Kim's test (Algorithm~\ref{algo:Kim}) with $\alpha = 2$ (i.e., collision entropy) provides better estimates than the compression estimator since the bias is reduced. However, for a BMS with $p=0.5$, Algorithm~\ref{algo:Kim} with $\alpha = 2$ is slightly worse than the compression estimator. It is because the bias is zero for BMS with $p=0.5$ (see Remark~\ref{rem:extreme}) and the variance of estimates increases for the higher $\alpha$. Fig.~\ref{fig:bms} shows that Algorithm~\ref{algo:Kim} with $\alpha=3$ suffers from larger variances for high entropy sources. Hence, we focus on $\alpha = 2$ for Algorithm~\ref{algo:Kim} since the variance is manageable and its computations are efficient. 
	
	Table~\ref{tab:bms} shows the mean squared error (MSE) and the mean percentage error (MPE) of all the estimators for the BMS. Suppose that the correct (actual) min-entropy is $h$ and the estimates are $\widehat{h}_n$ for $n \in \{1,\ldots,N\}$. Then, the MSE and MPE are defined as:
	\begin{align}
	\mathsf{MSE} &= \frac{1}{N}\sum_{n=1}^{N}(h - \widehat{h}_n)^2, \\
	\mathsf{MPE} &= \frac{\SI{100}{\%}}{N}\sum_{n=1}^{N}{\frac{h - \widehat{h}_n}{h}}.  
	\end{align}  
	The MPEs are used to capture the sign of the error, which is not captured by MSE~\cite{Kelsey2015predictive}. We observe that the proposed estimator based on the collision entropy (Algorithm~\ref{algo:Kim}) can improve the estimation accuracy compared to other estimators. 
		

	Fig.~\ref{fig:near_inv} compares the min-entropy estimators for near-uniform distributed sources and inverted near-uniform distributed sources. As shown in Fig.~\ref{fig:near_inv}\subref{fig:near}, all the estimators are accurate for near-uniform distributed sources since all the estimators perform their estimation tasks by assuming near-uniform distribution, i.e., the lower bound on the min-entropy is the same as the actual min-entropy. 
	
	\begin{figure}[t]
		\centering
		\includegraphics[width=0.4\textwidth]{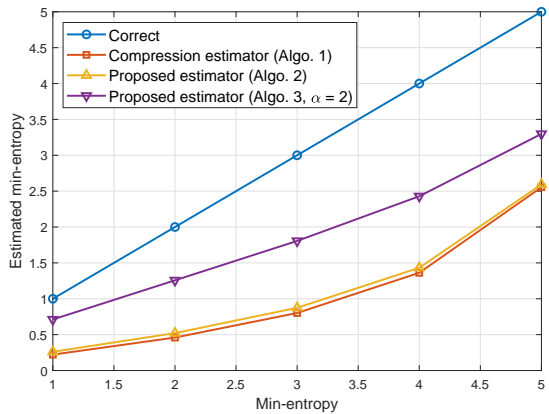}
		\caption{Comparison of min-entropy estimators for normal distributed sources.}
		\label{fig:normal}
	\end{figure}	
		
	On the other hand, the min-entropy estimates for inverted near-uniform distributed sources are quite underestimated as shown in Fig.~\ref{fig:near_inv}\subref{fig:inv}. The reason is that the inverted near-uniform distribution corresponds to the upper bounds in Fig.~\ref{fig:gap}, which leads to the maximal bias. The proposed estimator based on the collision entropy effectively reduces the bias, so it can provide much more accurate estimates than the other estimators. 
		
	Fig.~\ref{fig:normal} compares the min-entropy estimators for the normal distributed sources (rounded to integers). For this distribution, it is known that the compression estimator is prone to significant underestimates~\cite{Kelsey2015predictive}. The proposed estimator based on Coron's test is slightly better than the compression estimator. More importantly, the proposed estimator based on the collision entropy provides much more accurate estimates. 
		
	\begin{figure}[t]
		\centering
		\includegraphics[width=0.4\textwidth]{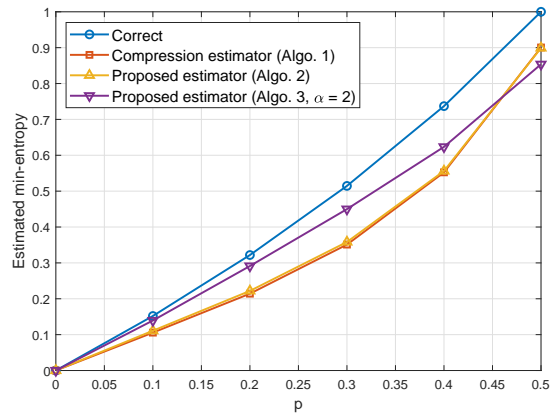}
		\caption{Comparison of min-entropy estimators for the first order Markov sources with $p = p(1|0) = p(0|1)$.}
		\label{fig:markov}
	\end{figure}

	Fig.~\ref{fig:markov} compares the min-entropy estimators for the first order Markov sources with $p = p(1|0) = p(0|1)$. The compression estimator and the proposed estimator based on Coron's test are almost identical. The proposed estimator based on the collision entropy is better than the other estimators since the bias is effectively reduced. However, at $H^{(\infty)}(\mathcal{S}) = 1$, its estimated min-entropy is lower than others because its variance is large as discussed in Section~\ref{sec:bias_var}. 
		
	\begin{table}[!t]
		\renewcommand{\arraystretch}{1.2}
		\caption{Per-bit Min-entropy Estimate for Real World Sources}
		\vspace{-2mm}
		\label{tab:real}
		\centering
		\begin{tabular}{|c|c|c|c|}	\hline
			& Compression estimator & Proposed & Proposed \\  
			& (Algo.~\ref{algo:nist}) & (Algo.~\ref{algo:Coron}) & (Algo.~\ref{algo:Kim})  \\ \hline \hline
			{RANDOM.ORG}  & 0.9110 & 0.9006 & 0.8690 \\ \hline
			{Ubld.it} & 0.8811 & 0.8811 & 0.8175  \\ \hline
			{LKRNG} & 0.9219 & 0.9006 & 0.8690 \\ \hline		  						
		\end{tabular}
		\vspace{-2mm}
	\end{table} 	
	
	We also evaluate min-entropy estimates using random number generators deployed in the real-world as in~\cite{Kelsey2015predictive}. The true entropies for these sources are unknown, so the MSE and MPE cannot be calculated. The min-entropy estimates of the real-world sources are presented in Table~\ref{tab:real}. We evaluate	RANDOM.ORG, Ubld.it, and Linux kernel random number generator (LKRNG). RANDOM.ORG~\cite{Random} is a service that provides random numbers based on atmospheric noise and Ubld.it generates random numbers by a TrueRNG device by~\cite{Ubld}. As we expected, the compression estimator (Algorithm~\ref{algo:nist}) and the proposed min-entropy estimator based on Coron's test (Algorithm~\ref{algo:Coron}) are almost identical. The proposed estimator based on the collision entropy (Algorithm~\ref{algo:Kim}) is slightly lower than the others. It is because these real-world sources are high entropy sources, which make the variance of estimates by Algorithm~\ref{algo:Kim} larger than the variance of other estimators as observed in Fig.~\ref{fig:bms} and Fig.~\ref{fig:markov}.

	\begin{table}[!t]
	\renewcommand{\arraystretch}{1.2}
	\caption{Execution Time for 100 Sample Sequences (Seconds)}
	\vspace{-2mm}
	\label{tab:time}
	\centering
	\begin{tabular}{|c|c|c|}	\hline
		Compression estimator & Proposed estimator & Proposed estimator \\ 
		(Algo.~\ref{algo:nist}) & (Algo.~\ref{algo:Coron}) & (Algo.~\ref{algo:Kim})  \\ \hline \hline
		9946.02 & 20.32 & 7.76 \\ \hline
	\end{tabular}
	\vspace{-2mm}
	\end{table} 	 	
	
	Table~\ref{tab:time} compares the execution times of estimators. We evaluate 100 sample sequences of 6,000,000 bits and set $M = 1,000$. We experiment on a Window 10 machine with an Intel i9-10900KF processor (3.7 GHz) and 32 GB of RAM. Although the execution times depend on the experimental environments, Table~\ref{tab:time} supports that the proposed algorithms are much more efficient than the compression estimator. 

\begin{figure}[t]
	\centering
	\subfloat[]{\includegraphics[width=0.4\textwidth]{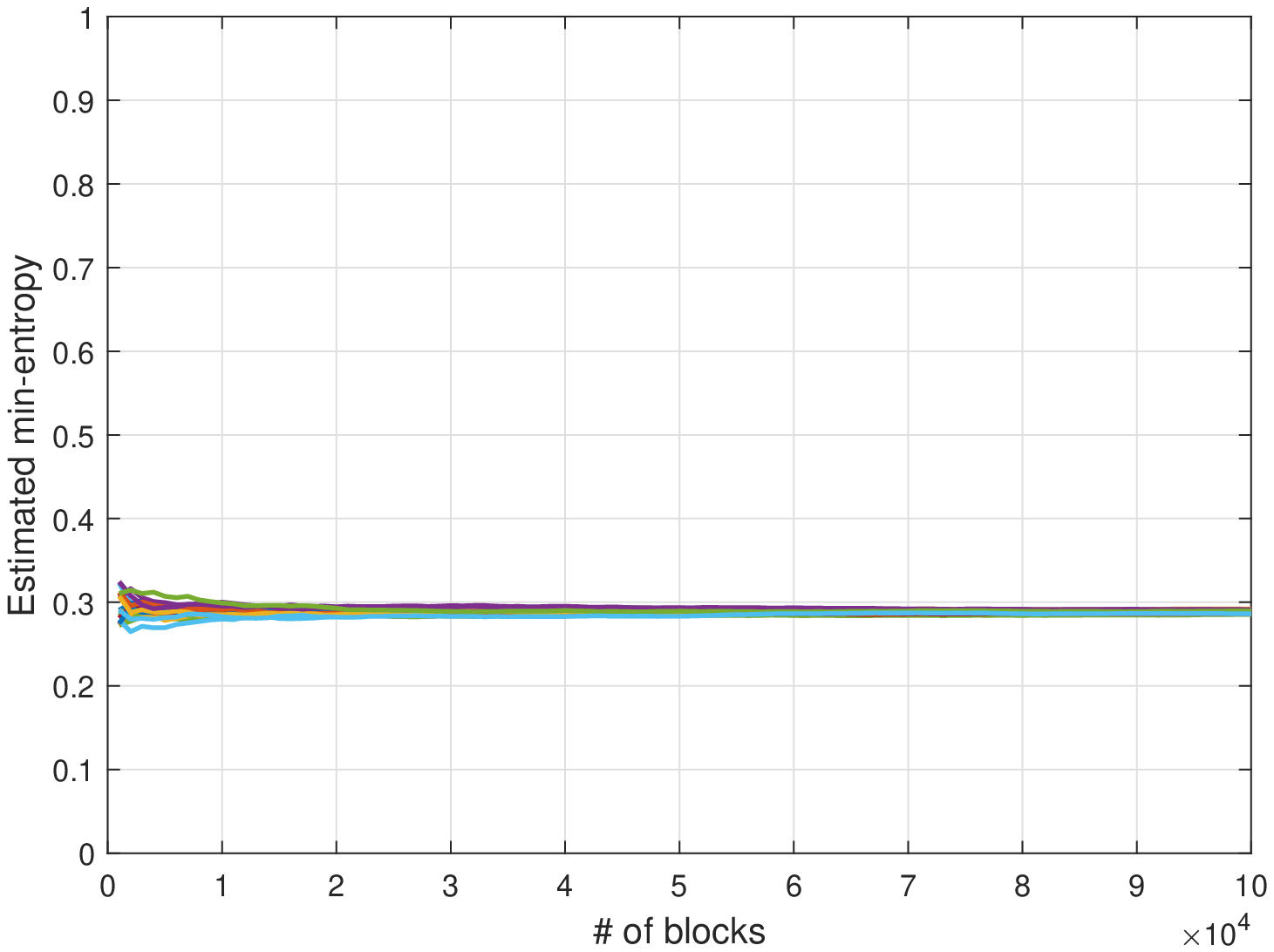}
		\label{fig:online_2}}
	\hfil
	\subfloat[]{\includegraphics[width=0.4\textwidth]{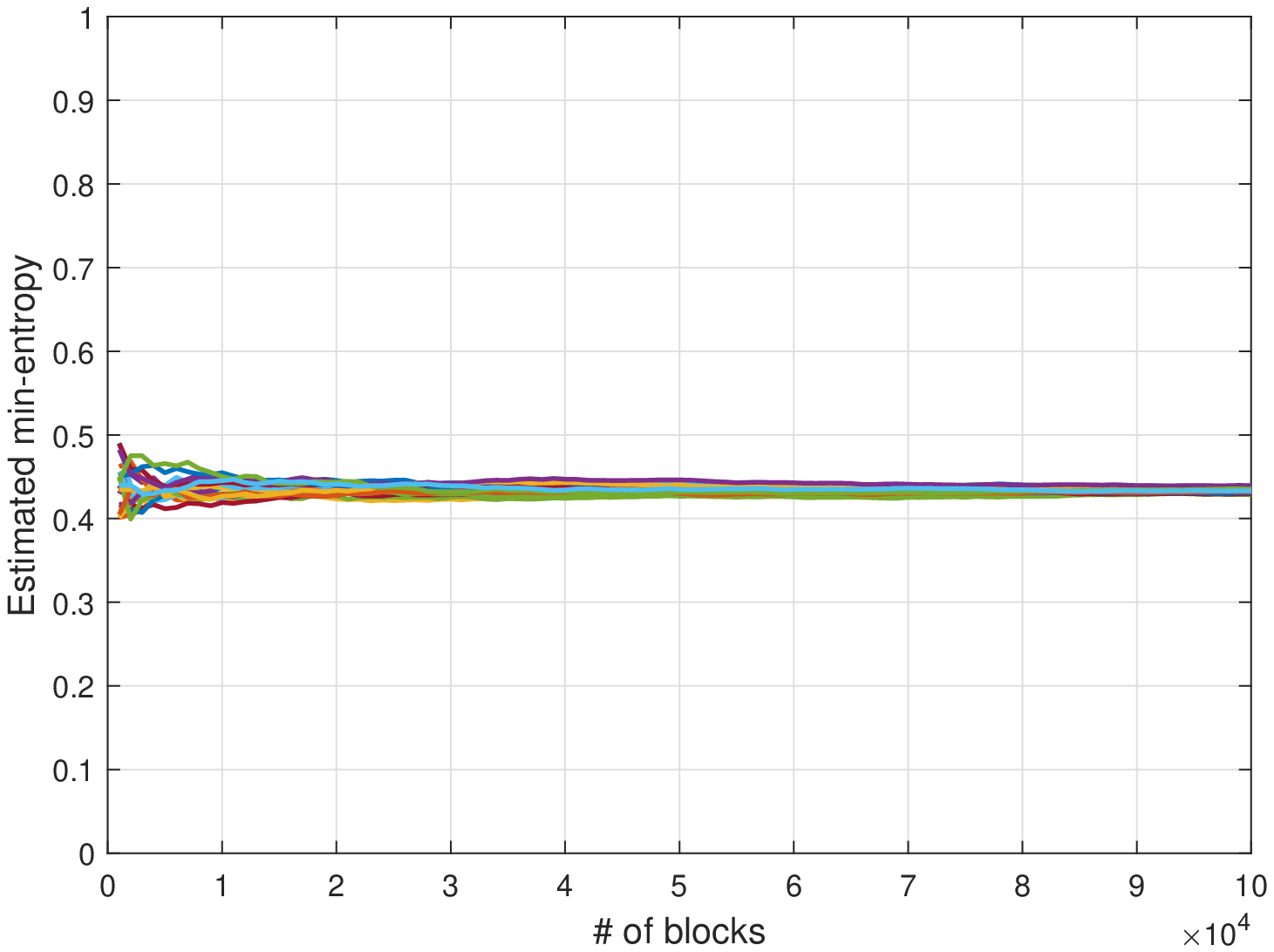} \label{fig:online_3}}
	\hfill
	\subfloat[]{\includegraphics[width=0.4\textwidth]{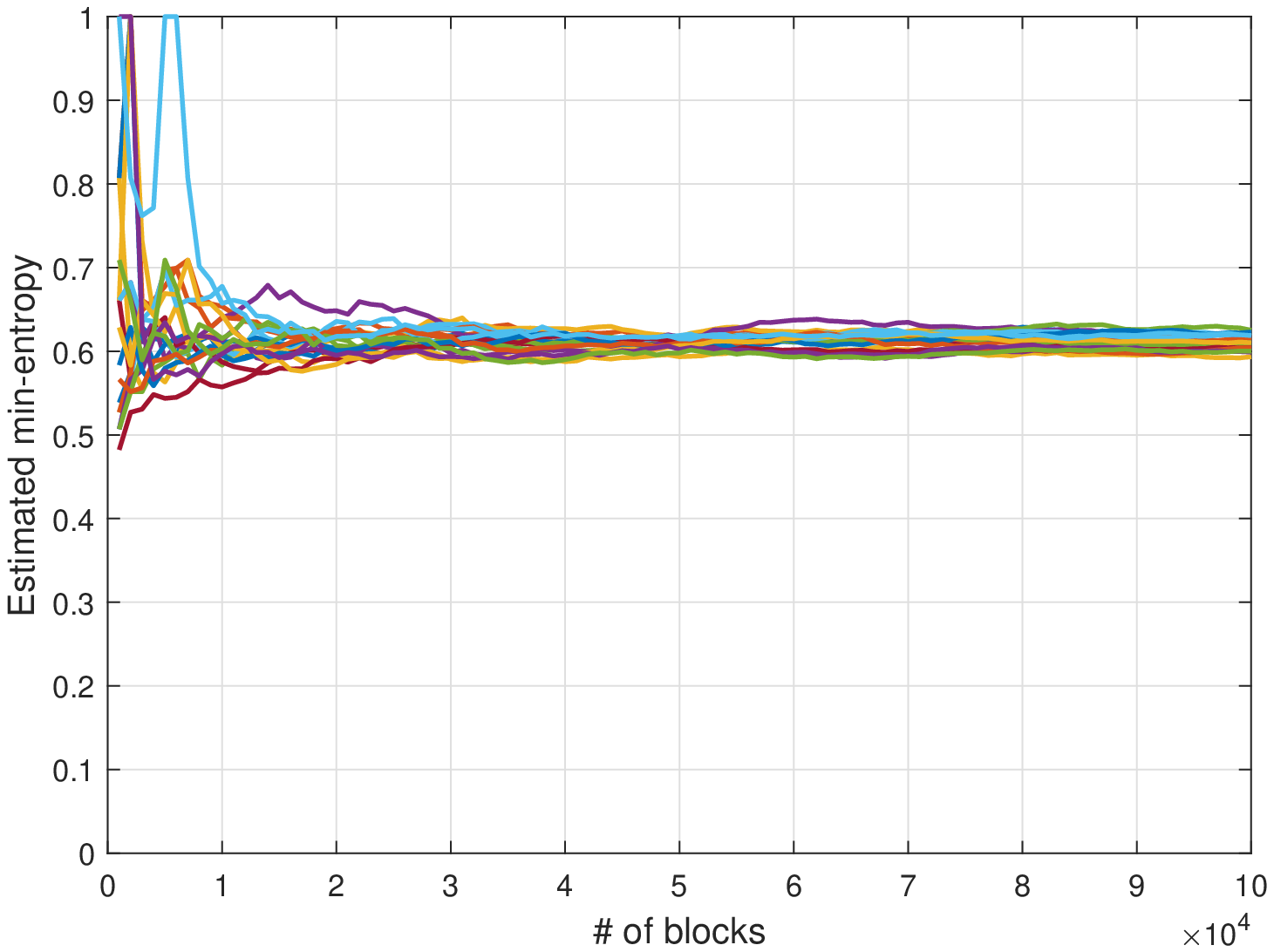}
	\label{fig:online_4}}
	\hfil
	\caption{Online min-entropy estimates by Algorithm~\ref{algo:online} for 20 BMS with $p$. Estimates of 20 sample sources are plotted and each curve corresponds to a sample source: (a) $p=0.2$, (b) $p=0.3$, and (c) $p=0.4$.}
	\label{fig:online}
	\end{figure}

	Fig.~\ref{fig:online} shows the min-entropy estimates by online algorithm (Algorithm~\ref{algo:online}) for BMS with $p$. Algorithm~\ref{algo:online} can output an estimate as getting a new block $b_k$. As collecting more samples, the estimate is improved and its variance is reduced. We observe that higher entropy sources result in larger variances as discussed in Section~\ref{sec:bias_var}. 
	
	We note that Algorithm~\ref{algo:online} is very effective to detect low entropy sources. It is because the proposed online estimator provides estimates as getting new blocks and low entropy sources can be detected reliably with limited samples. For example, Fig.~\ref{fig:online} shows that a low entropy source whose min-entropy is less than 0.5 can be detected by testing only several hundred blocks, which is comparable to the required samples ($>10\times 2^L$) for initialization of the compression estimator.

	\section{Conclusion}\label{sec:conclusion}

	We proposed computationally efficient min-entropy estimators by leveraging the variations of Maurer's test. The proposed estimator based on Coron's test achieves almost identical accuracy with much less computation than the compression estimator. Moreover, we propose the min-entropy estimator based on the collision entropy. It has advantages over the compression estimator in terms of estimation accuracy, computational complexity, and data efficiency at the cost of variance. We also propose a lightweight estimator which processes data samples in an online manner. The proposed online estimator can estimate the min-entropy with limited samples and then improve its estimation accuracy as getting more samples. Since the proposed online estimator does not need to store the entire samples, it is proper for applications with stringent resource constraints.
		
	\appendices
	
	\section{Proof of Theorem~\ref{thm:order_gap}} \label{pf:order_gap}
	We show that $\theta^{(\alpha)} \ge \theta^{(\alpha+1)}$ for $\theta^{(\alpha)} \gg \frac{1}{1 + (B-1)^{\frac{\alpha - 1}{\alpha}}}$, which is equivalent to \eqref{eq:order_improve}. For convenience, suppose that $x = \theta^{(\alpha)}$ and $y = \theta^{(\alpha + 1)}$. 
	
	In~\cite{Beck1990upper}, it was shown that $\frac{\alpha - 1}{\alpha}H^{(\alpha)} \le \frac{\beta - 1}{\beta}H^{(\beta)}$ for $\beta > \alpha$ and $\alpha\beta > 0$. If $\beta = \alpha + 1$ and $\alpha > 1$, 
	\begin{equation} \label{eq:Renyi_inequality}
	H^{(\alpha)}(\mathcal{B}) \le \frac{\alpha^2}{\alpha^2 - 1} H^{(\alpha+1)}(\mathcal{B})
	\end{equation}		
	Then, we obtain the following inequality for the near-uniform distribution:
	\begin{align} 
	& \frac{1}{1-\alpha}\log_2{\left(x^\alpha + \frac{(1-x)^\alpha}{(B-1)^{\alpha-1}}\right)} \nonumber \\
	& \le \frac{\alpha}{1 - \alpha^2} \log_2{\left(y^{\alpha+1} + \frac{(1 - y)^{\alpha + 1}}{(B-1)^{\alpha}}\right)}, \label{eq:Renyi_inequality_near}
	\end{align} 
	which is equivalent to
	\begin{align} 
	& \left(x^\alpha + \frac{(1-x)^\alpha}{(B-1)^{\alpha-1}}\right)^{\frac{1}{\alpha}} \nonumber \\
	& \ge \left(y^{\alpha+1} + \frac{(1 - y)^{\alpha + 1}}{(B-1)^{\alpha}}\right)^{\frac{1}{\alpha + 1}}.\label{eq:order_inequality} 
	\end{align}		
	If $x^\alpha \gg \frac{(1-x)^\alpha}{(B-1)^{\alpha-1}}$ and $y^{\alpha+1} \gg \frac{(1-y)^{\alpha+1}}{(B-1)^{\alpha}}$, then \eqref{eq:order_inequality} becomes $x \ge y$. Hence, $\theta^{(\alpha)} \ge \theta^{(\alpha+1)}$ for $\theta^{(\alpha)} \gg \frac{1}{1 + (B-1)^{\frac{\alpha - 1}{\alpha}}}$.

	\section{Proof of Theorem~\ref{thm:slope}}\label{pf:slope}
	
	Suppose that $\theta = \frac{1}{B} + \delta$ where $\delta \ll \frac{1}{B}$. Then, \eqref{eq:def_z} becomes
	\begin{align}
	z(\theta, \alpha) 
	&= \frac{1}{\alpha}\cdot \frac{1}{\left(\frac{1}{B}+\delta \right)^{\alpha-1} - \left( \frac{1 - \frac{1}{B}-\delta}{B-1}\right)^{\alpha-1}}   \\
	&= \frac{1}{\alpha} \cdot \frac{B^{\alpha-1}}{ \left(1 + B\delta \right)^{\alpha - 1} - \left(1 - \frac{B}{B-1}\delta \right)^{\alpha - 1} } \\
	&\simeq \frac{1}{\alpha}\cdot\frac{B^{\alpha-1}}{\{1+(\alpha-1)B\delta\} - \{1 - (\alpha-1)  \frac{B\delta}{B-1}\}} \label{eq:slope_approx} \\
	&=\frac{B^{\alpha-3}}{\alpha(\alpha - 1)}\cdot \frac{B-1}{\delta}. \label{eq:slope_pf}  
	\end{align}
	where \eqref{eq:slope_approx} follows from $(1+B\delta)^{\alpha-1}\simeq 1 + (\alpha-1)B\delta$ and $\left(1 - \frac{B}{B-1}\delta \right)^{\alpha-1} \simeq 1 - (\alpha - 1)\frac{B}{B-1}\delta$ for $\delta \ll \frac{1}{B}$. It is straightforward to derive \eqref{eq:xi} from \eqref{eq:slope_pf}.   

	\section{Proof of Lemma~\ref{lem:var_g}}\label{pf:var_g}
	
	For $\alpha = 2$,  $\mathbb{E}(g_{\mathcal{K}}(D, 2)) = P(D=1)$ by \eqref{eq:g_collision}. Also, 
	\begin{align}
		\mathbb{E}\left(g_{\mathcal{K}}(D, \alpha=2)^2 \right) &= \sum_{i=1}^{K}{P(D=i)g_{\mathcal{K}}(D=i, \alpha=2)^2} \nonumber \\
		&= P(D=1). 
	\end{align} 
	Then, 
	\begin{equation} \label{eq:var_pf}
		\mathsf{Var}(g_{\mathcal{K}}(\vect{s}, \alpha = 2)) = P(D=1) - P(D=1)^2.
	\end{equation}
	
	From~\eqref{eq:g_Kim}, we obtain
	\begin{equation} \label{eq:g_Kim3}
		g_{\mathcal{K}}(i, 3) = 
		\begin{cases}
			1, & \text{if } i = 1;\\
			-1, & \text{if } i = 2; \\
			0, & \text{otherwise}.  
		\end{cases}
	\end{equation}
	Then, we can derive $\mathbb{E}(g_{\mathcal{K}}(D,3)) = P(D=1) - P(D=2)$ and $	\mathbb{E}\left(g_{\mathcal{K}}(D,3)^2 \right) = P(D=1) + P(D=2)$. Hence, 
	\begin{align}
		& \mathsf{Var}(g_{\mathcal{K}}(\vect{s}, \alpha = 3)) \nonumber \\
		&= P(D=1) + P(D=2) - \{P(D=1) - P(D=2)\}^2 \nonumber \\
		& =  \mathsf{Var}(g_{\mathcal{K}}(\vect{s}, \alpha = 2)) + \{P(D=2) - P(D=2)^2\} \nonumber \\
		&\quad + 2P(D=1)P(D=2) \label{eq:var_ineq_0}\\
		&\ge \mathsf{Var}(g(D,\alpha=2)) \label{eq:var_ineq}
	\end{align} 
	where \eqref{eq:var_ineq_0} follows from \eqref{eq:var_pf} and \eqref{eq:var_ineq} follows from $P(D=2) \ge P(D=2)^2$ and $P(D)\ge 0$.
	
	\section{Proof of Theorem~\ref{thm:var}}\label{pf:var}
	
	By Lemma~\ref{lem:var_g}, $\mathsf{Var}(g_{\mathcal{K}}(D, 2)) \le \mathsf{Var}(g_{\mathcal{K}}(D, 3))$. Also, the corrective factor for $\alpha = 3$ is slightly greater than the corrective factor for $\alpha = 2$. We show that $z(\theta,2) < z(\theta,3)$ if $\theta < \frac{2}{3} - \frac{1}{3(B-2)}$. Then, $\mathsf{Var}(\theta^{(2)}) < \mathsf{Var}(\theta^{(3)})$ for $\theta < \frac{2}{3} - \frac{1}{3(B-2)}$.   
	
	From~\eqref{eq:def_z}, the inequality $z(\theta,\alpha) < z(\theta,\alpha+1)$ is equivalent to
	\begin{align} \label{eq:var_alpha}
	&(\alpha + 1) \left\{ \theta^\alpha - \left(\frac{1-\theta}{B-1}\right)^\alpha  \right\} \nonumber \\
	& < \alpha \left\{ \theta^{\alpha-1} - \left(\frac{1-\theta}{B-1}\right)^{\alpha-1}  \right\}.
	\end{align}
	For $\alpha = 2$, \eqref{eq:var_alpha} becomes 
	\begin{equation} \label{eq:var_alpha_2}
	\left(\theta - \frac{1}{B}\right)(3(B-2)\theta - 2B + 5)<0, 
	\end{equation}
	which is equivalent to $\frac{1}{B} < \theta < \frac{2B-5}{3(B-2)} = \frac{2}{3} - \frac{1}{3(B-2)}$. Note that $\frac{1}{B} < \frac{2}{3} - \frac{1}{3(B-2)}$ for $B > 3$, which holds $L \ge 2$. Since $\theta > \frac{1}{B}$ by definition, we obtain $z(\theta,\alpha) < z(\theta,\alpha+1)$ if $\theta < \frac{2}{3} - \frac{1}{3(B-2)}$.

	\section*{Acknowledgment}
	
	The authors would like to thank the anonymous reviewers for detailed suggestions that significantly improved the paper.

	
	\bibliographystyle{IEEEtran}
	\bibliography{abrv,mybib}

\end{document}